\documentclass[12pt,a4paper,final]{iopart}

\usepackage{iopams}  
\usepackage{cite}
\usepackage{graphicx}
\usepackage[breaklinks=true,colorlinks=true,linkcolor=blue,urlcolor=blue,citecolor=blue]{hyperref}

\begin{document}

\title[]{Quality of joint remote preparation of an arbitrary two-qubit state under the effect of noise}

\author{Thanh Dat Le and Van Hop Nguyen}
\address{Faculty of Physics, Hanoi National University of Education, 136 Xuan Thuy, Cau Giay, Hanoi, Vietnam}
\ead{hopnv@hnue.edu.vn}

\begin{abstract}
We address the issue of improving the quality of the joint remote preparation of an arbitrary two-qubit state in case four qubits of the quantum channel which consists of a GHZ state and a GHZ-like one are subjected to noises. Two controlling parameters are added, one in the quantum channel and other in the measurement of the second sender, in order to optimize the averaged fidelities. The results from analyzing the behaviors of the optimal averaged fidelities show that there are essentially two different ways for the optimization of the efficiency of the protocol. The first is simply choosing suitably the quantum channel as well as the measurement in which the desired fidelity can be found in large values of noisy parameters. The second is by means of interactions between qubits and dissipative environments whose result is more noises more fidelity.

\end{abstract}

\pacs{03.67.-a}
\vspace{2pc}
\noindent{\it Keywords}: joint remote state preparation, optimal averaged fidelity, quantum noise

\section{Introduction}

\indent{Joint remote state preparation (JRSP) \cite{i1,i2,i3} is one of the most interesting quantum transmission protocols in quantum information processing. In JRSP, several senders located in separated sites have a task to transmit a quantum state to a remote receiver via an entangled quantum channel shared beforehand among all the people in the protocol. The basic distinction between JRSP protocols and remote state preparation (RSP) \cite{i4} ones is that each of the senders in JRSP holds partially the classical information of the prepared state, so none of them can reveal the full. Since firstly introduced in \cite{i1}, JRSP has been received a great attention and investigated in different points of view \cite{i5, i6, i7, i8, i9, i10, i11, i12, i13, i14, i15, i16, i17, i18, i19, i20, i21, i22}. It has shown an advancement in JRSP protocols, by employing suitable measurement schemes in which the senders implement their measurements depending upon the measurement results of the previous senders, JRSP became deterministic \cite{i10, i16, i18, i19}. Besides, the experimental architecture of JRSP protocol has been put forward \cite{i23} and an approach to perform JRSP of photonic states with linear optical devices has been recently studied \cite{i24}.}\\ 
\indent{In realistic quantum communication processing the presence of noise which is essentially the interactions with surrounding environments is unavoidable. The consequence of noise is usually to decrease entanglement of the quantum channel and therefore lead to the reduction of the quality of the protocol. To deal with such difficulty the first solution is that via legitimate procedures the noisy channel is transformed into a better one. In this connection, there are two possible ways being proposed, namely, quantum distillation \cite{i25,i26,i27} which destroys some noisy entangled pairs to create the one with desired entanglement and weak measurement \cite{i29,i30} using non-unitary operators to protect the quantum channel. The drawback of both techniques is the success probability being less than unity. Several studies related to the improvement of the quantum teleportation protocol under the effect of noise have exploited quantum distillation \cite{i25,i31} and weak measurement \cite{i32,i33}. However, there is another solution suggesting that instead of transforming the noisy channel the stages of the protocol are modified in an appropriate way to achieve a maximum transmission fidelity. Applying this approach to the noisy quantum teleportation has been studied in the literature \cite{i34,i35,i36,i37,i38,i39}.}\\ 
\indent{JRSP protocols in the noisy scenarios have also been investigated through solving the Lindblad master equations \cite{i40,i41} or using Kraus operators \cite{i42,i43,i44}. However, these papers have just showed the dependence of the fidelity of the protocol on parameters of noise or the quantum channel and none of them uses the techniques quantum distillation, weak measurement or modification of the protocol stages. Recently, JRSP of a qubit in the presence of noise in which the initial quantum channel and the steps of the protocol are suitably chosen to optimize the fidelity has been put forward \cite{i45}. Like in Ref. \cite{i45}, in this paper, the same issue is addressed but for the case of a two-qubit. Particularly, we make use of Kraus operators to take into account the joint remote state preparation of a two-qubit state in the presence of four typical noisy channels, namely, the bit-flip, phase-flip, depolarizing and amplitude-damping channel \cite{i46,i47}. By means of adjustment in the standard JRSP protocol, the averaged fidelities are optimized and then analyzed through their phase diagrams. The results show that the protocol is more robust with respect to the amplitude-damping or phase-flip noise than the other noises as the optimal averaged fidelity exceeding the classical limit in case of qubits suffering such noises is found in a larger domain of noise parameters. In case the environment noise is bit-flip the second sender, who produces the quantum channel, can apply the Pauli operator $X$ to obtain the desired fidelity at a large value of noise parameter. Some specific scenarios, in addition, show that less quantum entanglement or greater noisy strength parameters can heighten the quality of JRSP protocol. From these results, we categorize more precisely two ways for the optimization of the protocol according to their features.}\\ 
\indent{This paper is outlined as follows. In Sec. 2, we take a brief view of JRSP of an arbitrary two-qubit state in density operators representation. We then optimize the values of the averaged fidelities obtained in various scenarios of noises and analyze their phase diagrams in Sec. 3. Finally, Sec. 4 is devoted to conclusions.}

\section{JRSP of an arbitrary two-qubit state in density operators representation}
\indent{Suppose that Alice and Bob wish to help Charlie remotely prepare a two-qubit state in the following form} 
\begin{equation}\label{psi}
\left| \psi  \right\rangle =\lambda_0\left| 00 \right\rangle + \lambda_1 e^{i\varphi_1}\left| 01 \right\rangle +\lambda_2e^{\varphi_2}\left| 10 \right\rangle + \lambda_3e^{i\varphi_3}\left| 11 \right\rangle,
\end{equation}
in which $\varphi_i\, (i=\overline {1,3})$ and $\lambda_j\,(j=\overline{0,3})$ are real parameters and
\begin{equation}
\lambda_0^2+\lambda_1^2+\lambda_2^2+\lambda_3^2=1.
\end{equation}
For simplicity, we denote $\left| \textbf{0} \right\rangle_{rs}= \left| 00 \right\rangle_{rs}, \left |\textbf{1} \right\rangle = \left| 01 \right\rangle_{rs}, \left|\textbf{2} \right\rangle_{rs}=\left | 10\right\rangle_{rs}$ and $\left |\textbf{3} \right\rangle_{rs} = \left |11 \right\rangle_{rs}$.\\
The classical information of the state $\left| \psi\right\rangle $ is divided between Alice and Bob in such a way that  Alice holds information about amplitude $\left\{ \lambda_0,\lambda_1,\lambda_2,\lambda_3 \right\}$ and Bob holds information about phase $\left\{\varphi_1, \varphi_2, \varphi_3 \right\}$. To jointly prepare a two-qubit state the quantum channel is at least made up of six qubits in which qubits 1 and 2, qubits 3 and 4 and qubits 5 and 6 belong to Alice, Bob and Charlie, respectively. Therefore, in density language it can be denoted as $\rho_{123456}$. The most general JRSP of a two-qubit state contains three steps as follows:\\
\textbf{Step 1}: Alice measures qubits 1 and 2 in the basis $\{\left|\omega_{k}\right\rangle_{12};k=\overline {0,3}\}$,
\begin{equation}\label{UA}
\left|\omega_{k} \right\rangle_{12} = \sum\limits_{l=0}^{3} {a_{kl}(\lambda_0,\lambda_1,\lambda_2,\lambda_3)\left|\textbf{l} \right\rangle_{12}},
\end{equation}
where $a_{kl}(\lambda_0,\lambda_1,\lambda_2,\lambda_3)$ are coefficients which depend on $\lambda_0,\lambda_1,\lambda_2,\lambda_3$. Right after obtaining the outcome $\left| \omega_k\right\rangle_{12}$, she uses two classical bits to public $k$ and the state of the quantum channel reduces into an entangled state connecting Bob and Charlie
\begin{equation}
\rho^{(k)}_{3456}=\frac{_{12}\left\langle \omega_{k}\right|\rho_{123456}\left|\omega_{k}\right\rangle_{12}}{P^{(k)}},
\end{equation}
with
\begin{equation}
P^{(k)}=Tr\left(_{12}\left\langle \omega_{k}\right|\rho_{123456}\left|\omega_{k}\right\rangle_{12}\right)
\end{equation}
is the probability that the measurement result of Alice is $k$. \\
\textbf{Step 2}: Based on the value of $k$, Bob measures his qubits 3 and 4 in the basis $\Big\{\left|\sigma^{(k)}_{m}\right\rangle_{34};m=\overline{0,3}\Big\}$,
\begin{equation}\label{V}
\left| \sigma^{(k)}_{m}\right\rangle_{34}=\sum\limits_{n=0}^{3} {b_{mn}}(\varphi_1, \varphi_2,\varphi_3,k) \left| \textbf{n} \right\rangle_{34},
\end{equation}
where $b_{mn}(\varphi_1,\varphi_2,\varphi_3,k)$ are coefficients depending on $\varphi_1,\varphi_2,\varphi_3, k$. If Bob's result is $\left|\sigma^{(k)}_{m}\right\rangle_{34}$, $m$ is publicly broadcast (of course, by two classical bits) and the state $\rho^{(k)}_{3456}$ transforms into
\begin{equation}
\rho^{(km)}_{56}=\frac{_{34}\left\langle \sigma^{(k)}_{m}\right| \rho^{(k)}_{3456}\left|\sigma^{(k)}_{m}\right\rangle_{34}}{P^{(km)}},
\end{equation}
in which
\begin{equation}
P^{(km)}=Tr\left( _{34}\left\langle \sigma^{(k)}_{m}\right| \rho^{(k)}_{3456}\left|\sigma^{(k)}_{m}\right\rangle_{34} \right)
\end{equation}
is the probability of Bob's outcome of $m$.\\
\textbf{Step 3}: Finally, according to the values of $k$ and $m$ announced by Alice and Bob, Charlie applies to $\rho^{(km)}_{56}$ an appropriate unitary operator $R^{(km)}$ to reconstruct the desired state
\begin{equation}
\widetilde \rho^{(km)}=R^{(km)}\rho^{(km)}_{56}\left[ R^{(km)} \right]^{\dag }.
\end{equation}
The degree of closeness between $\widetilde \rho ^{(km)}$ and the transmitted state $\left| \psi \right\rangle$ in Eq. (\ref{psi}), fidelity of the protocol, is quantified by 
\begin{equation}
F^{(km)}=\left\langle \psi \right| \widetilde \rho^{(km)} \left|\psi \right\rangle,
\end{equation}
and is averaged over all possible measurement results
\begin{equation}
F=\sum\limits_{k=0}^{3}\sum\limits_{m=0}^{3} P^{(k)}P^{(km)}F^{(km)}.
\end{equation}
In order to have the fidelity being independent of the prepared state the amplitude parameters of the input state should be reparameterised
\begin{eqnarray}
\left\{\begin{array}{l}
\lambda_{0} = \cos{\eta_3},\\
\lambda_{1} = \sin{\eta_3}.\cos{\eta_2},\\
\lambda_{2} = \sin{\eta_3}.\sin{\eta_2}.\cos{\eta_1},\\
\lambda_{3} = \sin{\eta_3}.\sin{\eta_2}.\sin{\eta_1}.  
\end{array}\right.
\end{eqnarray}
Then, with the assumption of a uniform distribution, the ultimate averaged fidelity $\left\langle F \right\rangle $ can be calculated in the following \cite{i48}
\begin{equation}
\fl \left\langle F \right\rangle=\frac{3!}{\pi^3}\int\limits_{0}^{\frac{\pi}{2}}{d\eta_1} \int\limits_{0}^{\frac{\pi}{2}}d\eta_2\int\limits_{0}^{\frac{\pi}{2}}d\eta_3\int\limits_{0}^{2\pi}d\varphi_1 \int\limits_{0}^{2\pi}d\varphi_2 \int\limits_{0}^{2\pi}d\varphi_3 \prod\limits_{i=1}^{3}{F.\cos{\eta_i}.\left(\sin{\eta_i}\right)^{2i-1}}.
\end{equation}

\section{JRSP of an arbitrary two-qubit state under the effect of noise}
Firstly, we consider the perfect JRSP of an arbitrary two-qubit state in noiseless environment. The quantum channel being made use of is a product state of two maximally entangled Greenberger-Horne-Zeilinger (GHZ) states
\begin{equation}
\left| Q \right\rangle_{135246}= \frac{1}{\sqrt{2}}\left( \left|000\right\rangle+\left|111\right\rangle\right)_{135}\otimes \frac{1}{\sqrt{2}}\left( \left|000\right\rangle+\left|111\right\rangle\right)_{246}.\label{Qi}
\end{equation}
The coefficients in Eqs. (\ref{UA}) and (\ref{V}) are chosen and displayed as the elements of the following unitary matrices
\begin{eqnarray}
a_{kl}(\lambda_0,\lambda_1,\lambda_2,\lambda_3)=A_{kl}(\lambda_0,\lambda_1,\lambda_2,\lambda_3),\\
A(\lambda_0,\lambda_1,\lambda_2,\lambda_3)=\left( {\begin{array}{*{20}{c}}
\lambda_0&-\lambda_1&-\lambda_2&-\lambda_3\\
\lambda_1&\lambda_0&\lambda_3&-\lambda_2\\
\lambda_2&-\lambda_3&\lambda_0&\lambda_1\\
\lambda_3&\lambda_2&-\lambda_1&\lambda_0
\end{array}} \right),\label{Ue}
\end{eqnarray}
and
\begin{eqnarray}
b_{mn}(\varphi_1,\varphi_2,\varphi_3,k)=B_{mn}(\varphi_1,\varphi_2,\varphi_3,k),\\
B(\varphi_1,\varphi_2,\varphi_3,k)= T^{(k)}.B(\varphi_1, \varphi_2,\varphi_3),\label{Ve}
\end{eqnarray}
where
\begin{eqnarray}
T^{(k)}= (-iY)^{[k/2]} \otimes Z^{[k/2]}(-iY)^{k\,mod\,2},\label{T}\\
B(\varphi_1, \varphi_2,\varphi_3)=\frac{1}{2}\left(
\begin{array}{cccc}
 1 & 1 & 1 & 1 \\
 e^{-i \varphi_1} & -e^{-i \varphi_1} & e^{-i \varphi_1} & -e^{-i \varphi_1} \\
 e^{-i \varphi_2} & e^{-i \varphi_2} & -e^{-i \varphi_2} & -e^{-i \varphi_2} \\
 e^{-i \varphi_3} & -e^{-i \varphi_3} & -e^{-i \varphi_3} & e^{-i \varphi_3} \\
\end{array}
\right).
\end{eqnarray}
Then
\begin{equation}
R^{(km)}=Z^{[m/2]}X^{[k/2]} \otimes Z^{m\,mod\,2}X^{k\,mod\,2}. \label{Rklmn}
\end{equation}
Note that $I$ is the $2\times2$ identity matrix, $X, Y$ and $Z$ are the standard Pauli matrices and $[ \bullet]$ denotes the floor function. Correspondingly, $F^{(km)}=1$ for any $k,m,\lambda_i\, (i=\overline {0,3} )$ and $\varphi_{j}\,(j=\overline {1,3})$, which means not only the averaged fidelity but also the success probability is unit. Thus, in this case we obtain a perfect two-qubit JRSP.\\ 
\indent{In noisy case, the quantum channel is chosen as follows}
\begin{equation}
\left|Q(\theta)\right\rangle_{135246} =\left| Q\right\rangle_{135}\otimes \left|Q(\theta) \right\rangle_{246},\label{Qcn}
\end{equation}
in which 
\begin{eqnarray}
\left| Q \right\rangle_{135} &=\frac{1}{\sqrt{2} }(\left|000\right\rangle+\left|111\right\rangle)_{135},\\
 \left|Q(\theta) \right\rangle_{246} &= (\cos{\theta}\left|000\right\rangle+\sin{\theta}\left|111\right\rangle)_{246}.\label{Q246}
\end{eqnarray}
The matrice $A(\lambda_0,\lambda_1,\lambda_2,\lambda_3)$ chosen in Eq. (\ref{Ue}) is kept unchanged but the one $B(\varphi_1,\varphi_2,\varphi_3,k)$ in Eq. (\ref{Ve}) is replaced by
\begin{eqnarray}\label{Vnew}
 B(\varphi_1,\varphi_2,\varphi_3,k,\xi) = T^{(k)}.B(\varphi_1,\varphi_2,\varphi_3,\xi),
\end{eqnarray}
 where
\begin{eqnarray}
 B(\varphi_1,\varphi_2,\varphi_3,\xi)=\nonumber \\
 \frac{1}{{\sqrt{2}}}\left(
\begin{array}{cccc}
 \cos (\xi ) & \cos (\xi ) & \sin (\xi ) & \sin (\xi ) \\
 e^{-i \varphi_1} \cos (\xi ) & -e^{-i \varphi_1} \cos (\xi ) & e^{-i \varphi_1} \sin (\xi ) & -e^{-i \varphi_1} \sin (\xi ) \\
 e^{-i \varphi_2} \sin (\xi ) & e^{-i \varphi_2} \sin (\xi ) & -e^{-i \varphi_2} \cos (\xi ) & -e^{-i \varphi_2} \cos (\xi ) \\
 e^{-i \varphi_3} \sin (\xi ) & -e^{-i \varphi_3} \sin (\xi ) & -e^{-i \varphi_3} \cos (\xi ) & e^{-i \varphi_3} \cos (\xi ) \\
\end{array}
\right).
\end{eqnarray}
Note that $\theta$ and $\xi$, respectively, in Eqs. (\ref{Qcn}) and (\ref{Vnew}), are the free controlling parameters for the sake of the optimization of the JRSP protocol.\\ 
In this paper, we deal with four typical types of noise, namely, the bit-flip (B), phase-flip (P), amplitude-damping (A) and depolarizing (D). These noises can be expressed in terms of Kraus operators \cite{i46}
\begin{eqnarray}
K^{(B)}_1(p_{B})=\sqrt{1-p_{B}}\,I,\,K^{(B)}_2(p_{B})=\sqrt{p_{B}}\,X,\\
K^{(P)}_1(p_{P})=\sqrt{1-p_{P}}\,I,\, K^{(P)}_2(p_{P})=\sqrt{p_{nP}}\,Z,\\
K^{(A)}_1(p_{A})=\left( {\begin{array}{*{20}{c}}
{1}&{0}\\
{0}&{\sqrt{1-p_{A}}}
\end{array}} \right),\,
K^{(A)}_2(p_{A})=\left( {\begin{array}{*{20}{c}}
{0}&{\sqrt{p_{A}}}\\
{0}&{0}
\end{array}} \right)
\end{eqnarray} 
and
\begin{eqnarray}
K^{(D)}_1(p_{D})=\sqrt{1-\frac{3}{4}p_{D}}\,I,\, K^{(D)}_2(p_{D})=\sqrt{\frac{1}{4}p_{D}}\,X, \nonumber \\
K^{(D)}_3(p_{D})=\sqrt{\frac{1}{4}p_{D}}\,Y,\,K^{(D)}_4(p_{D})=\sqrt{\frac{1}{4}p_{D}}\,Z.
\end{eqnarray}
Suppose that each of the six qubits 1, 2, 3, 4, 5 and 6 independently suffers a type of noise then the influence of noises is modeled by virtue of superoperator that takes the initial quantum channel $\left|Q(\theta)\right\rangle_{135246}$ into a mixed state in the following linear map
\begin{eqnarray}
\fl \rho^{(\alpha\gamma\epsilon\beta\delta\zeta)}_{135246} &= & \sum\limits_{i=1}^{N_\alpha} \sum\limits_{k=1}^{N_\gamma} \sum\limits_{m=1}^{N_\epsilon}K^{(\alpha)}_i(p_{1\alpha}) \otimes K^{(\gamma)}_{k}(p_{3\gamma}) \otimes K^{(\epsilon)}_m(p_{5\epsilon}). \left| Q\right\rangle_{135}\left\langle Q\right|.\Big[K^{(\alpha)}_i(p_{1\alpha}) \otimes K^{(\gamma)}_{k} \nonumber \\
\fl &&  (p_{3\gamma}) \otimes K^{(\epsilon)}_m(p_{5\epsilon})\Big]^{\dag} \otimes \sum\limits_{j=1}^{N_\beta}\sum\limits_{\ell=1}^{N_\delta}\sum\limits_{n=1}^{N_\zeta} K^{(\beta)}_{j}(p_{2\beta})\otimes K^{(\delta)}_{\ell}(p_{4\delta}) \otimes K^{(\zeta)}_{n}(p_{6\zeta}).\left| Q(\theta)\right\rangle_{246}\nonumber\\
\fl && \left\langle Q(\theta)\right|.\Big[K^{(\beta)}_{j}(p_{2\beta})\otimes K^{(\delta)}_{\ell}(p_{4\delta}) \otimes K^{(\zeta)}_{n}(p_{6\zeta}) \Big]^{\dag},
\end{eqnarray}
in which $K^{(\alpha)}_j(p_{1\alpha})$ and $p_{1\alpha}\, (0\leq p_{1\alpha} \leq 1)$ are the $j^{th}$ Kraus operator and the noise strength of the noisy channel $\alpha \in \{B, P, A, D \}$ that affects qubit 1 and $N_\alpha$ is the number of $\alpha$-type noise Kraus operators. There holds the same explanations for $K^{(\beta)}_j(p_{2\beta}), K^{(\gamma)}_k(p_{3\gamma}), K^{(\delta)}_{\ell}(p_{4\delta}), K^{(\epsilon)}_{m}(p_{5\epsilon}), K^{(\zeta)}_n(p_{6\zeta}), p_{2\beta}, p_{3\gamma},p_{4\delta},p_{5\epsilon},p_{6\zeta}$ and $N_\beta, N_\gamma, N_\delta, N_\epsilon, N_\zeta$.\\
\indent{The overall noisy scenario we concern is in the following. Let Bob be the producer who first produces the quantum channel at his site. Afterwards, he sends qubits 1 and 2 through similar $\alpha-$type noisy channels to Alice as well as qubits 5 and 6 through similar $\gamma-$type noisy channels to Charlie, but keeps qubits 3 and 4 with himself. In general, the noise strength is a parameterized quantity which is proportional to the time the noise is acting on the qubit or the distance the qubit has to travel along in the noisy environment. Thus, we can assume that $p_{1\alpha}=p_{2\alpha}=p_{a\alpha}$ and $p_{5\gamma}=p_{6\gamma}=p_{c\gamma}$ and the quantum channel becomes 
\begin{eqnarray}
\fl \rho^{(\alpha\gamma)}_{135246} &=& \sum\limits_{i=1}^{N_\alpha}\sum\limits_{k=1}^{N_\gamma} K^{(\alpha)}_i(p_{a\alpha}) \otimes I\otimes K^{(\gamma)}_k(p_{c\gamma}).\left|Q\right\rangle_{135}\left\langle Q \right|. \left[K^{(\alpha)}_i(p_{a\alpha}) \otimes I\otimes K^{(\gamma)}_k(p_{c\gamma}) \right]^{\dag}\nonumber \\
\fl &&\otimes \sum\limits_{j=1}^{N_\alpha}\sum\limits_{\ell=1}^{N_\gamma} K^{(\alpha)}_{j}(p_{a\alpha})\otimes I \otimes K^{(\gamma)}_{\ell}(p_{c\gamma}).\left| Q(\theta)\right\rangle_{246}\left\langle Q (\theta)\right|.\Big[K^{(\alpha)}_{j}(p_{a\alpha})\otimes I \otimes K^{(\gamma)}_{\ell} \nonumber \\
\fl && (p_{c\gamma}) \Big]^{\dag}.
\end{eqnarray}
To begin, address the situation in which $\alpha = B$ and $\gamma \in \{B, P, A, D\}$. Following the steps of JRSP of a two-qubit state in presence of noise and with the notation of $\left\langle F _{B \gamma}\right\rangle\, (\gamma \in \{B,P,A,D\})$ as the averaged fidelities corresponding to the present case, one obtains
\begin{eqnarray}
\fl \left\langle F_{BB} \right\rangle &=& \frac{2}{5}+\frac{1}{5} \left[p_{{aB}} \left(2 p_{{cB}}-1\right)-p_{{cB}}\right] \left[p_{{aB}} \left(2 p_{{cB}}-1\right)-p_{{cB}}+2\right]+ \frac{1}{80} \big[(\pi ^2-16) p_{{aB}}
\nonumber\\
\fl && +16\big] \left(p_{{cB}}-1\right)^2 \big[\sin (2 \xi )+\sin(2\theta)\big]+\frac{1}{40}\left[8-(16-\pi ^2) (p_{{aB}}-p_{{aB}}^2)\right](p_{{cB}}\nonumber \\
\fl && -1)^2 \sin (2 \xi ) \sin (2 \theta) \label{FBB},\\
\fl \left\langle F_{BP}\right\rangle &=& \frac{2}{5}+\frac{1}{5} \left(p_{{aB}}-2\right) p_{{aB}}+\frac{1}{80} \left[(\pi ^2-16) p_{{aB}}+16\right] \left(1-2 p_{{cP}}\right)\big[ \sin (2 \xi)+\sin(2\theta)\big] \nonumber \\
\fl && +\frac{1}{40} \big[8-(16-\pi ^2)(p_{{aB}} - p_{{aB}}^2)\big]\left(2 p_{{cP}}-1\right)^2 \sin (2 \xi) \sin(2\theta),
\label{FBP}\\
\fl \left\langle F_{BA}\right\rangle &=&\frac{2}{5}+\frac{1}{20} \left[2 p_{{aB}} \left(p_{{cA}}-1\right)-p_{{cA}}\right] \left[2 p_{{aB}} \left(p_{{cA}}-1\right)-p_{{cA}}+4\right]+\frac{1}{20} \left(1-2 p_{{aB}}\right) p_{{cA}} \nonumber \\
\fl && \times \left[2 p_{{aB}} \left(p_{{cA}}-1\right)-p_{{cA}}+2\right]  \cos (2 \theta )+\frac{1}{160} \left[(\pi ^2-16) p_{{aB}}+16\right]\sqrt{1-p_{{cA}}} (2 \nonumber \\
\fl &&-p_{{cA}}) \big[ \sin (2 \xi)+\sin(2\theta)\big]+\frac{1}{160} \big[(\pi ^2-16) p_{{aB}}+16\big] \sqrt{1-p_{{cA}}} p_{{cA}}  \sin (2 \xi )\cos ( \nonumber \\
\fl && 2 \theta)+ \frac{1}{40} \big[8 -(16-\pi ^2) (p_{{aB}}-p_{{aB}}^2)\big]  \left(1-p_{{cA}}\right)\sin (2 \xi ) \sin (2 \theta ), \label{FBA}
\end{eqnarray}
and 
\begin{eqnarray}
\fl \left\langle F_{BD}\right\rangle &=& \frac{2}{5}+\frac{1}{20} \left[2 p_{{aB}} \left(p_{{cD}}-1\right)-p_{{cD}}\right] \left[2 p_{{aB}} \left(p_{{cD}}-1\right)-p_{{cD}}+4\right]+\frac{1}{160} \big[(\pi ^2-16) p_{{aB}} \nonumber \\
\fl && +16\big] \left(p_{{cD}}-2\right) \left(p_{{cD}}-1\right) \big[\sin (2 \xi )+\sin(2\theta)\big]+\frac{1}{40} \left[8-(16-\pi ^2) (p_{{aB}}-p_{{aB}}^2)\right] \nonumber \\
\fl && \times (1 -p_{{cD}})^2 \sin (2 \xi ) \sin (2 \theta ).\label{FBD}
\end{eqnarray}
Then, the parameters $\theta$ and $\xi$ are used for the optimization of the averaged fidelities. From Eqs. (\ref{FBB}) and (\ref{FBD}) and the notation in which $0 \le p_{a\alpha},\,p_{c\gamma}\le 1 $ for any $\alpha,\,\gamma \in \{B,P,A,D\}$, the expression as the function of $p_{aB}$ and $p_{cB}$ or $p_{aB}$ and $p_{cD}$ placing in the left side of $\big[\sin(2\xi)+\sin(2\theta)\big]$ or $\sin(2\xi)\sin(2\theta)$ is completely positive. Therefore, the optimal values of $\theta$ and $\xi$ that maximize $\left\langle F_{BB}\right\rangle $ and $\left\langle F_{BD} \right\rangle $ are
\begin{equation}
 \theta^{(BB)}_{opt}=\xi^{(BB)}_{opt}=\theta^{(BD)}_{opt}=\xi^{(BD)}_{opt}=\frac{\pi}{4}.
\end{equation}
In Eq. (\ref{FBP}), since the expression standing in front of $\sin(2\xi)\sin(2\theta)$ is always greater than zero and the signs of that in front of $\big[\sin(2\xi)+\sin(2\theta)\big]$ in case $p_{cP}<1/2$ and in case $p_{cP}>1/2$ are reversed the optimization for $\left\langle F_{BP}\right\rangle$ leads to
\begin{eqnarray}
\theta^{(BP)}_{opt}=\xi^{(BP)}_{opt}=\cases{\pi/4 &for $p_{cP} <1/2$,\\
-\pi/4 &for $p_{cP}>1/2$.\\}
\label{optconBP}
\end{eqnarray}
The remaining case of $\left\langle F_{BA} \right\rangle$ in Eq. (\ref{FBA}) is much more complicated. In spite of an easily-realized $\xi^{(BA)}_{opt}$ of $\pi/4$ the optimal value $\theta^{(BA)}_{opt}$ is determined from the equation $\partial \left\langle F_{BA} \right \rangle /\partial\, {\theta}|_{\theta=\theta^{(BA)}_{opt}}=0 $ and the condition $\partial^2 \left\langle F_{BA} \right\rangle/\partial^2\theta|_{\theta=\theta^{(BA)}_{opt}}<0 $.  Solving this equation with the condition one obtains
\begin{eqnarray}\label{tBA1}
\fl \theta_{opt}^{(BA)} = \frac{1}{2}\arctan \nonumber \\
\fl \frac{4 \left[8-(16-\pi ^2) \left(1-p_{{aB}}\right) p_{{aB}}\right] \left(1-p_{{cA}}\right)+\left[16-(16-\pi ^2) p_{{aB}}\right] \sqrt{1-p_{{cA}}} \left(2-p_{{cA}}\right)}{\big[(\pi ^2-16) p_{{aB}}+16\big] \sqrt{1-p_{{cA}}} p_{{cA}}+8 \left(1-2 p_{{aB}}\right) p_{{cA}} \big[-2 p_{{aB}} \left(1-p_{{cA}}\right)-p_{{cA}}+2\big]}
\end{eqnarray}
with $\sin{2\theta_{opt}^{(BA)}}>0$ and $\cos{2\theta_{opt}^{(BA)}}>0$ for $(p_{aB},p_{cA})$ satisfying the inequality (obviously $0\leq p_{aB},p_{cA}\leq1$)
\begin{equation}
\fl \big[(\pi ^2-16) p_{{aB}}+16\big] \sqrt{1-p_{{cA}}} p_{{cA}}+8 \left(1-2 p_{{aB}}\right) p_{{cA}} \big[-2 p_{{aB}} \left(1-p_{{cA}}\right)-p_{{cA}}+2\big]>0
\end{equation}
or with $\sin{2\theta_{opt}^{(BA)}}>0$ and $\cos{2\theta_{opt}^{(BA)}}<0$ for $(p_{aB},p_{cA})$ satisfying the inequality
\begin{equation}
\fl \big[(\pi ^2-16) p_{{aB}}+16\big] \sqrt{1-p_{{cA}}} p_{{cA}}+8 \left(1-2 p_{{aB}}\right) p_{{cA}} \big[-2 p_{{aB}} \left(1-p_{{cA}}\right)-p_{{cA}}+2\big]<0.
\end{equation}
With the values of $\theta^{(B\gamma)}_{opt}$ and $\xi^{(B\gamma)}_{opt}$ it is no difficulty to calculate the optimal averaged fidelities $\left\langle F_{B\gamma}\right\rangle_{opt}$ whose analytical expressions are
\begin{eqnarray}
\fl \left\langle F_{BB}\right\rangle_{opt} &=& \frac{2}{5}+\frac{1}{40}\Big\{2 p_{{aB}} \left(p_{{cB}}-1\right) \big[(\pi ^2-32) p_{{cB}}-\pi ^2+24\big]+8 \left(4 p_{{cB}}^2-8 p_{{cB}}+3\right) \nonumber \\
\fl && + p_{{aB}}^2 \big[(48-\pi^2) p_{{cB}}^2-2 (32-\pi^2) p_{{cB}}+24-\pi^2\big] \Big\},\\
\fl \left\langle F_{BP}\right\rangle_{opt} &=& \frac{2}{5}+\frac{1}{40} \Big\{\left[(\pi ^2-16) p_{{aB}}+16\right] \left| 1-2 p_{{cP}}\right| -\left[(\pi ^2-16) \left(p_{{aB}}-1\right) p_{{aB}}-8\right] \nonumber \\
\fl && \times\left(1-2 p_{{cP}}\right)^2+8 p_{{aB}}^2-16 p_{{aB}}\Big\},\label{FBPo}\\
\fl \left\langle F_{BD}\right\rangle_{opt} &=& \frac{2}{5}+\frac{1}{80}\Big\{48-2 (\pi ^2-24) p_{{aB}}^2 \left(p_{{cD}}-1\right){}^2+12 p_{{cD}} \left(3 p_{{cD}}-8\right)+p_{{aB}} \left(p_{{cD}}-1\right) \nonumber \\
\fl && \times \big[(3 \pi ^2-64) p_{{cD}}-4 \pi ^2+96\big]\Big\},
\end{eqnarray}
and
\begin{eqnarray}
\fl \left\langle F_{BA}\right\rangle_{opt} &=&\frac{2}{5}+\frac{1}{160} \Big\{\big[(16-\pi^2) p_{{aB}}-16\big] \sqrt{1-p_{{cA}}} \left(p_{{cA}}-2\right)+8\big[2 p_{{aB}} \left(p_{{cA}}-1\right)-p_{{cA}}\big] \nonumber \\
\fl && \times \big[2 p_{{aB}} \left(p_{{cA}}-1\right)-p_{{cA}}+4\big] \Big\}+\frac{1}{160}\Big\{\Big\{4 \left[8-(16-\pi ^2) \left(1-p_{{aB}}\right) p_{{aB}}\right] (1 \nonumber \\
\fl && -p_{{cA}}) + \big[16 -(16-\pi ^2) p_{{aB}}\big] \sqrt{1-p_{{cA}}} (2-p_{{cA}})\Big\}^2+ \Big\{\left[16-(16-\pi ^2) p_{{aB}}\right] \nonumber \\
\fl && \times \sqrt{1-p_{{cA}}} p_{{cA}}+8 (1 - 2 p_{{aB}}) p_{{cA}} \left[2 p_{{aB}} \left(p_{{cA}}-1\right)-p_{{cA}}+2\right]\Big\}^2\Big\}^{1/2}.
\end{eqnarray}
In Fig. (\ref{fgB}), the density plots of $\left\langle F_{B\gamma}\right\rangle_{opt}$ in corresponding $p_{aB}-p_{c\gamma} \, (\gamma\in\{B,P,A,D\})$ spaces are exploited to display the domain in which the protocol is useful. It deserves to emphasize that the requirement of the usefulness we address here means the optimal averaged fidelity of the JRSP protocol of a two-qubit state must exceed 2/5, the classical limit \cite{i49}.
\begin{figure}[ht!]
\begin{center}
\includegraphics[scale=0.55]{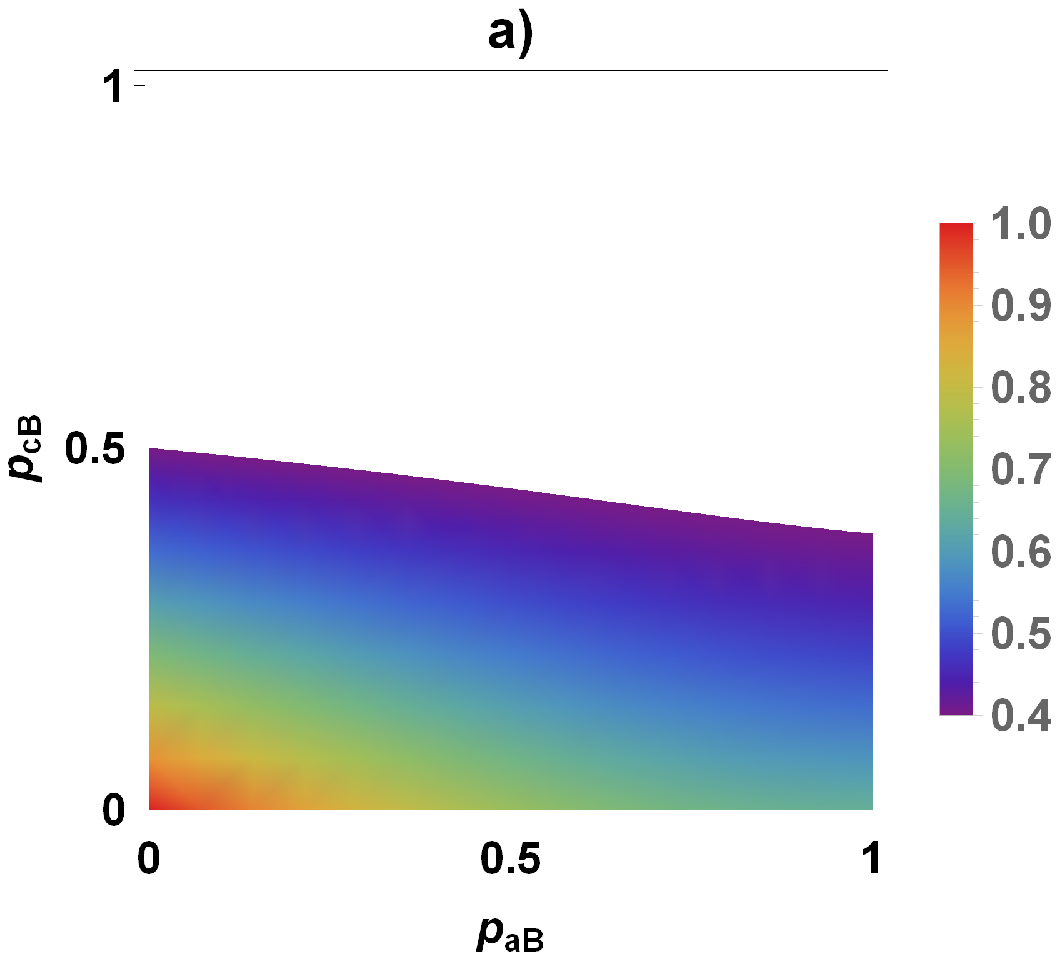}\quad\quad
\includegraphics[scale=0.55]{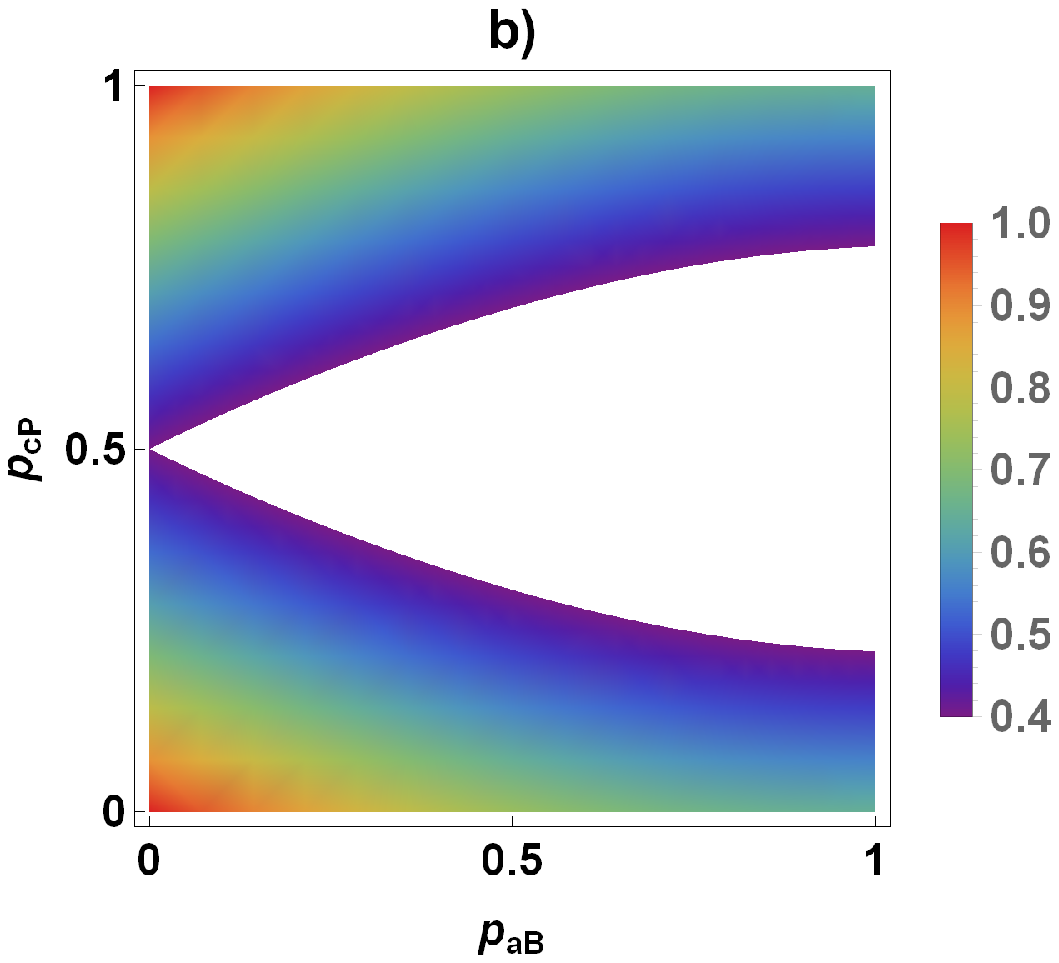}\vspace{0.2cm}
\includegraphics[scale=0.55]{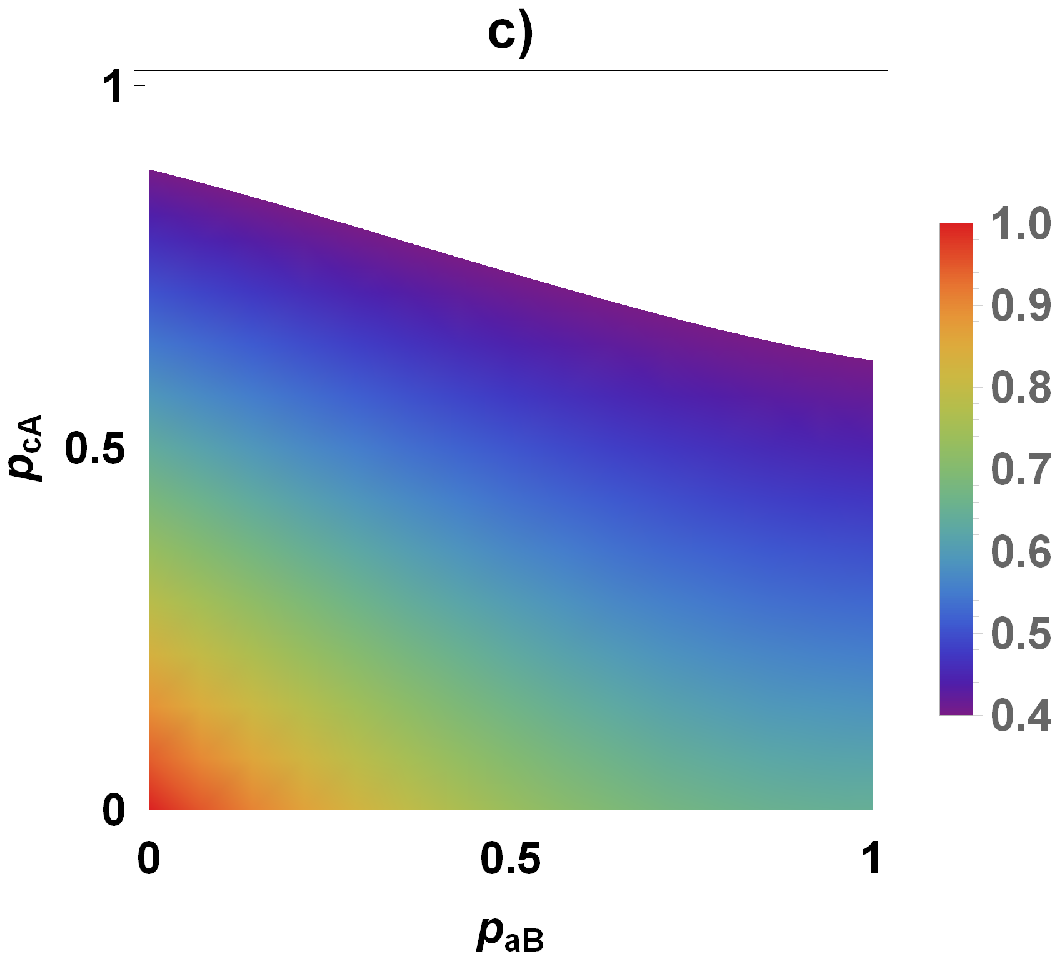}\quad\quad
\includegraphics[scale=0.55]{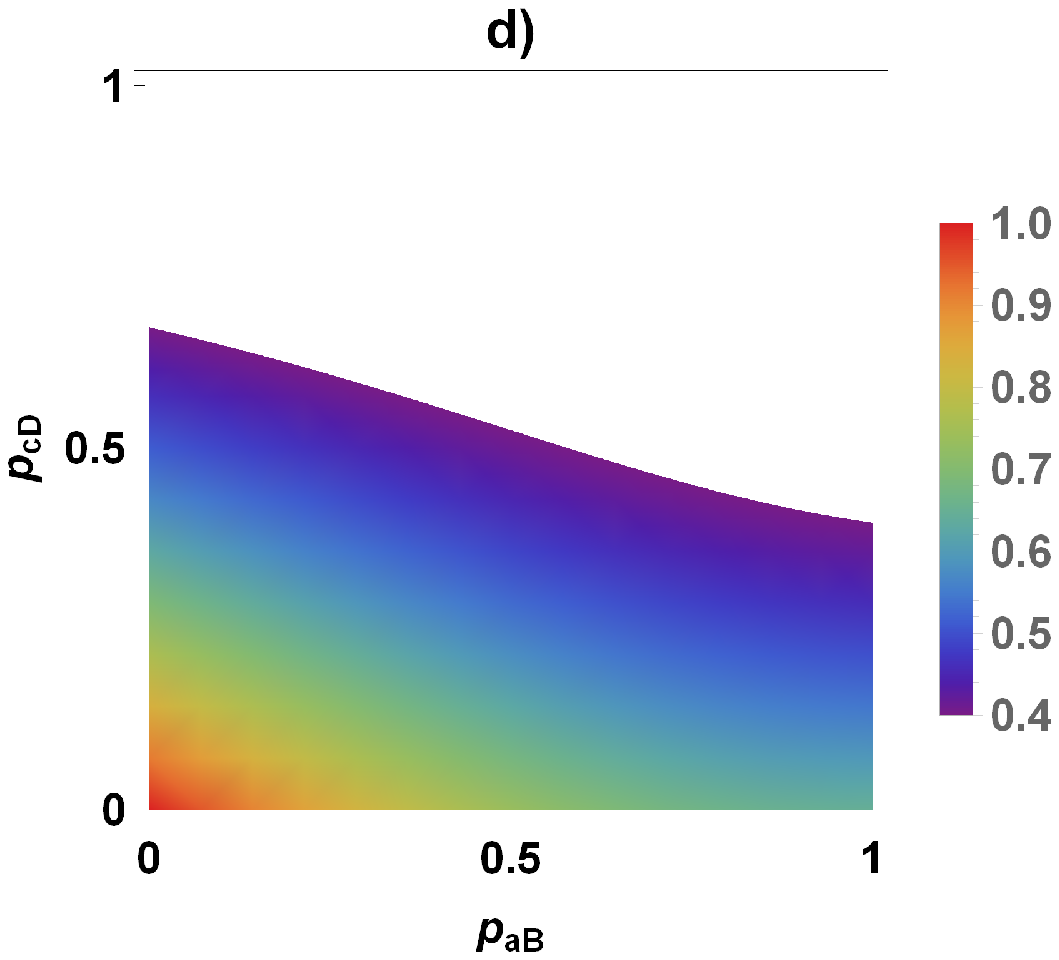}
\end{center}\vspace{-0.4cm}
\caption{Phase diagrams of the optimal averaged fidelities a) $\left\langle F_{BB}\right\rangle_{opt}$, b) $\left\langle F_{BP}\right\rangle_{opt}$, c) $\left\langle F_{BA}\right\rangle_{opt}$, and d) $\left\langle F_{BD}\right\rangle_{opt}$ in the $p_{aB}-p_{c\gamma}$ spaces. Colors illustrate the quantum values of $\left\langle F_{B\gamma}\right\rangle_{opt}$ and white background shows the classical domain.}
\label{fgB}
\end{figure}\\
Roughly speaking, from Figs. 1a, 1c and 1d, an increase in $p_{aB}$ or/and $p_{c\gamma}$ leads to a decrease in $\left\langle F_{B\gamma}\right\rangle _{opt}$, which shows that with a given bit-flip noise acting on qubits 1 and 2, no matter the bit-flip, amplitude-damping or depolarizing noise is added to be the sending environments of qubits 5 and 6 the quality of protocol will become poorer. For any $0\leq p_{aB}\leq 1$ in such plots there is always a chance to obtain the optimal averaged fidelity $\left\langle F_{B\gamma}\right\rangle_{opt}$ in quantum domain (i.e. the area in which $\left\langle F_{B\gamma}\right\rangle_{opt}>2/5$), while there shows limits of values of $p_{c\gamma}$, noted as $p_{c\gamma}^{lim}$, from which for any $p_{c\gamma}\geq p_{c\gamma}^{lim}$ the protocol is no longer useful. It can be understood that a greater value of $p_{c\gamma}^{lim}$ is equivalent to a weaker influence of $\gamma-$type noise on the protocol. Comparing three diagrams 1a, 1c, and 1d in more depth, one can easily see that $p_{cA}^{lim}>p_{cD}^{lim}>p_{cB}^{lim}$ and the area of the quantum domain in case $\gamma=A$ is the biggest and that in case $\gamma=B$ is the smallest. Different from the quantum domains of $\left\langle F_{B\gamma}\right\rangle_{opt}\, \left(\gamma \in \{B,A,D\}\right)$, the one of $\left\langle F_{BP}\right\rangle_{opt}$ in Fig. 1b is symmetric with respect to the segment $p_{cP}=1/2$, which results in the facts that a non-classical fidelity can be obtained even in the region containing large noise parameters. Such symmetry was found in Refs. \cite{i39, i45}, however, in this context it can be clearly shown from Eq. (\ref{FBPo}) in which $\left\langle F_{BP}\right\rangle_{opt}(p_{aB},1/2-\Delta p_{cP})=\left\langle F_{BP} \right\rangle_{opt}(p_{aB},1/2+\Delta p_{cP})$ $(0\leq\Delta p_{cP}\leq1/2)$ and its physical origin can be explained as follows.\\
Since the effects of noises are independent it has no loss of generality and is simply to consider the scenario in which qubits 1 and 2 aren't subjected to noises, but qubits 5 and 6 at the same time are affected by the phase-flip noise with noisy parameter $p_{P}$. It's necessary to recall the action of phase-flip on a qubit which is to flip phase of the qubit being in the excited state with the probability of $p_{P}$ and let the ground state unchanged with the probability of $1-p_{P}$. For convenience, let's denote $\rho^{+}_{135\,(246)}=\left| GHZ^{+} \right\rangle_{135\,(246)}\left\langle GHZ^{+} \right|  $ and $\rho^{-}_{135\,(246)}=\left| GHZ^{-} \right\rangle_{135\,(246)}\left\langle GHZ^{-} \right|$ with $\left|GHZ^{ \pm}\right\rangle=1/\sqrt{2}\big(\left|000\right\rangle \pm \left|111\right\rangle \big)$. In case $p_{P}$ is smaller than $1/2$, according to Eq. (\ref{optconBP}), the value of $\theta$ in Eq. (\ref{Q246}) is chosen as $\pi/4$. Then after being subjected to the phase-flip noise the initial quantum channel becomes a mixed state: $\left|Q(\frac{\pi}{4}) \right\rangle_{135246} \left\langle Q(\frac{\pi}{4}) \right| \to (1-p_P)^2\rho^{+}_{135} \otimes \rho^{+}_{246} + p_P(1-p_P)\rho^{+}_{135}\otimes \rho^{-}_{246}+p_P(1-p_P)\rho^{-}_{135}\otimes \rho^{+}_{246}+p_P^2 \rho^{-}_{135} \otimes \rho^{-}_{246} $, implying that if $p_P$ reduces to $0$, the after-subjected-to-noise quantum channel will be more similar to $\rho^{+}_{135} \otimes \rho^{+}_{246}$, which is equivalent to the quantum channel in noiseless case (Eq. (\ref{Qi})). Therefore, it can be seen that with $\theta = \pi/4$ and $\xi=\pi/4$ (note that $B(\varphi_1,\varphi_2,\varphi_3,k,\frac{\pi}{4})$ from Eq. (\ref{Vnew}) is the same to $B(\varphi_1,\varphi_2,\varphi_3,k)$ in Eq. (\ref{Ve})), the smaller value of $p_P$ is the closer to perfect JRSP this case is. Next, in case $p_P$ is larger than $1/2$, according to Eq. (\ref{optconBP}), $\theta$ in Eq. (\ref{Q246}) is given as $-\pi/4$. Similar to preceding case, the effect of the phase-flip noise is to transform the pure quantum channel into a mixed state:  $\left|Q(\frac{-\pi}{4}) \right\rangle_{135246} \left\langle Q(\frac{-\pi}{4}) \right| \to p_P(1-p_P)\rho^{+}_{135} \otimes \rho^{+}_{246} + (1-p_P)^2\rho^{+}_{135}\otimes \rho^{-}_{246}+p_P^2\rho^{-}_{135}\otimes \rho^{+}_{246}+p_P(1-p_P) \rho^{-}_{135} \otimes \rho^{-}_{246} $. So, it's clear that larger $p_{P}$ leads to the fact that the quantum channel under the effect of noises is closer to the state $\rho^{-}_{135}\otimes \rho^{+}_{246}$. However, one can check that in noiseless case, with the quantum channel chosen as $\left|GHZ^{-} \right\rangle_{135} \otimes \left|GHZ^{+} \right\rangle_{246} $, that is identical to the state $\rho^{-}_{135}\otimes \rho^{+}_{246}$, the matrices in Eqs. (\ref{Ue}) and (\ref{Rklmn}) being unchanged and the matrix of Eq. (\ref{Ve}) replaced by $B(\varphi_1,\varphi_2,\varphi_3,k,-\frac{\pi}{4})$ in Eq. (\ref{Vnew}), the JRSP protocol is perfect. As the result, it can be said that with both $\theta$ and $\xi$ chosen as $-\pi/4$, the larger $p_P$ is the closer to the perfect JRSP the present case is.\\
Motivated from the above explanation, by repeating calculations it's not that complicated to check that in order to obtain a quantum averaged fidelity even in the large range of the bit-flip noise strength Bob should first apply the Pauli operator $X$ to qubits before sending them via bit-flip environments. The results of this scheme being illustrated in Fig. (\ref{fgBL}) show that all the averaged fidelities amount to 1 at $(p_{aB}, p_{cB})=(1,1)$ or $(p_{aB}, p_{c\gamma})=(1,0)\, (\gamma \neq B)$. Hence, it is evident that different from the results of Refs. \cite{i39, i45} in case of bit-flip noise, a possible scheme in our paper can raise the fidelity when the noisy strength is considerable. It can be said that deciding whether the Pauli operator is applied before transmitting qubits can be understood as a kind of optimization.\\
In addition, with suitable selection of $p_{aB}$ and $p_{cA}$, the value of $\theta_{opt}^{(BA)}$ in Eq. (\ref{tBA1}) is different from $\pi/4$ at which the entangled state of qubits 2, 4 and 6 becomes a maximally entangled GHZ state, implying a better quality of the JRSP protocol in case of less entanglement. This result, that is to say, was also obtained in quantum teleportation \cite{i39} and JRSP of a single-qubit \cite{i45}. \\
\begin{figure}[ht!]
\begin{center}
\includegraphics[scale=0.55]{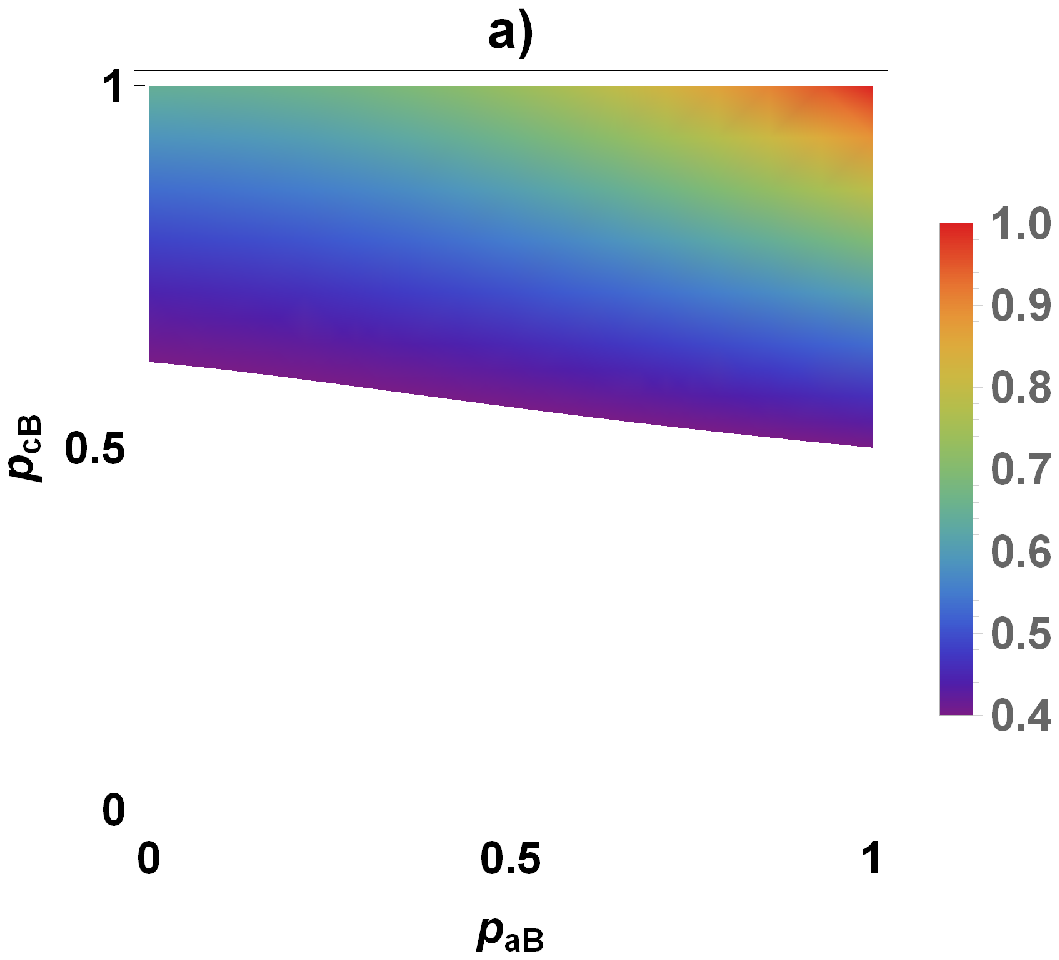}\quad\quad
\includegraphics[scale=0.55]{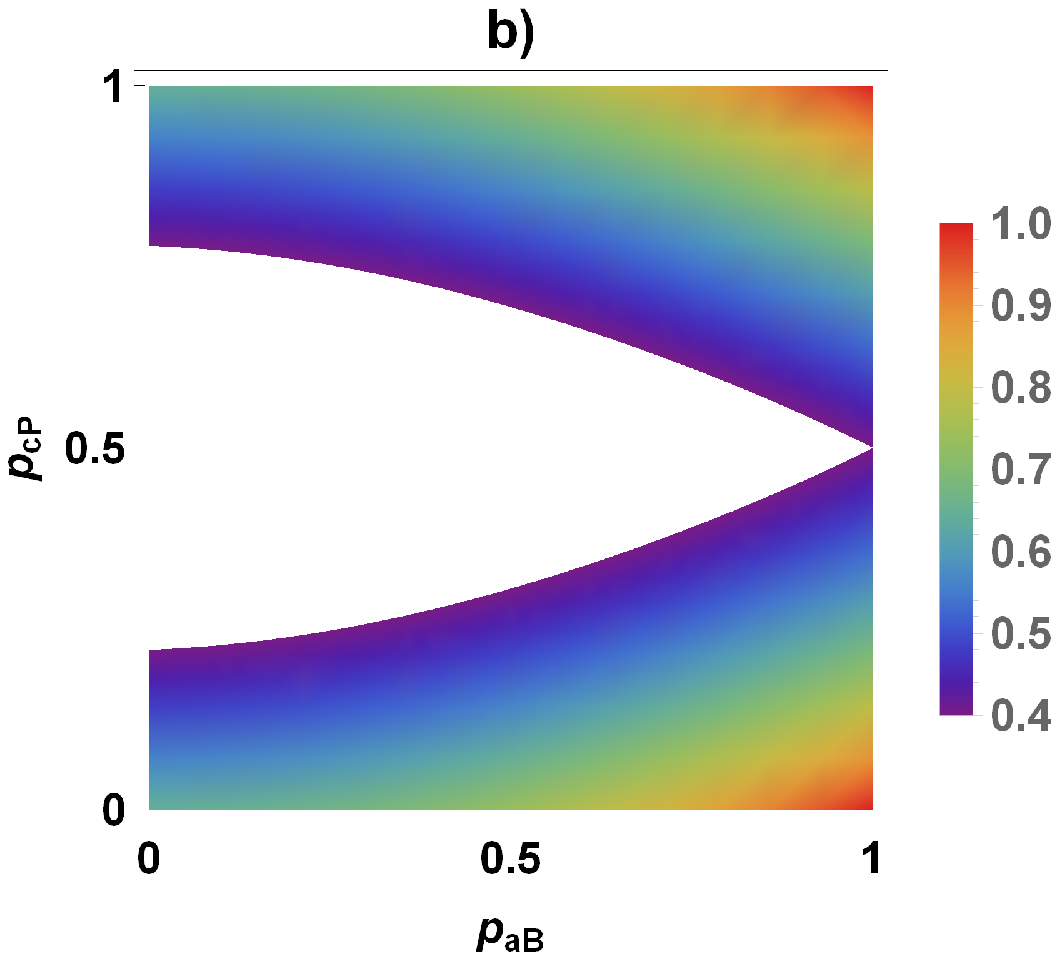}\vspace{0.2cm}
\includegraphics[scale=0.55]{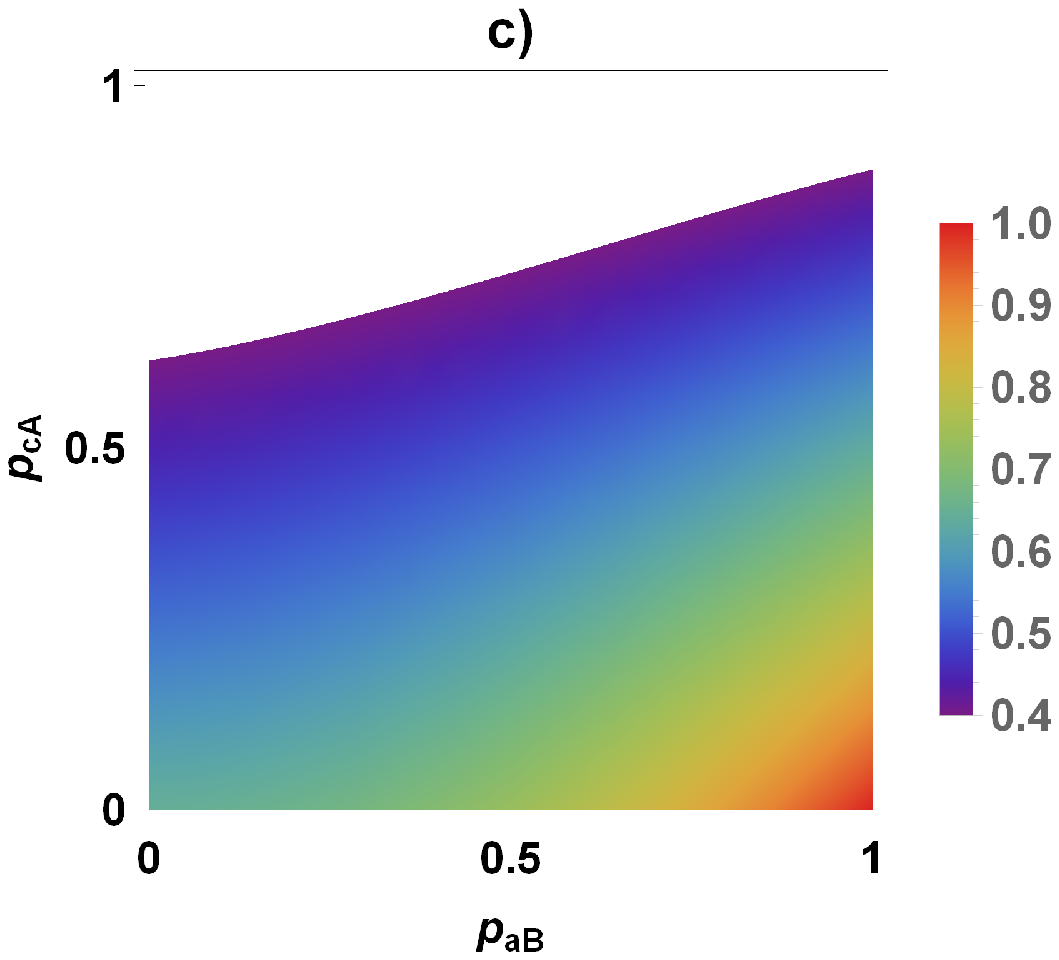}\quad\quad
\includegraphics[scale=0.55]{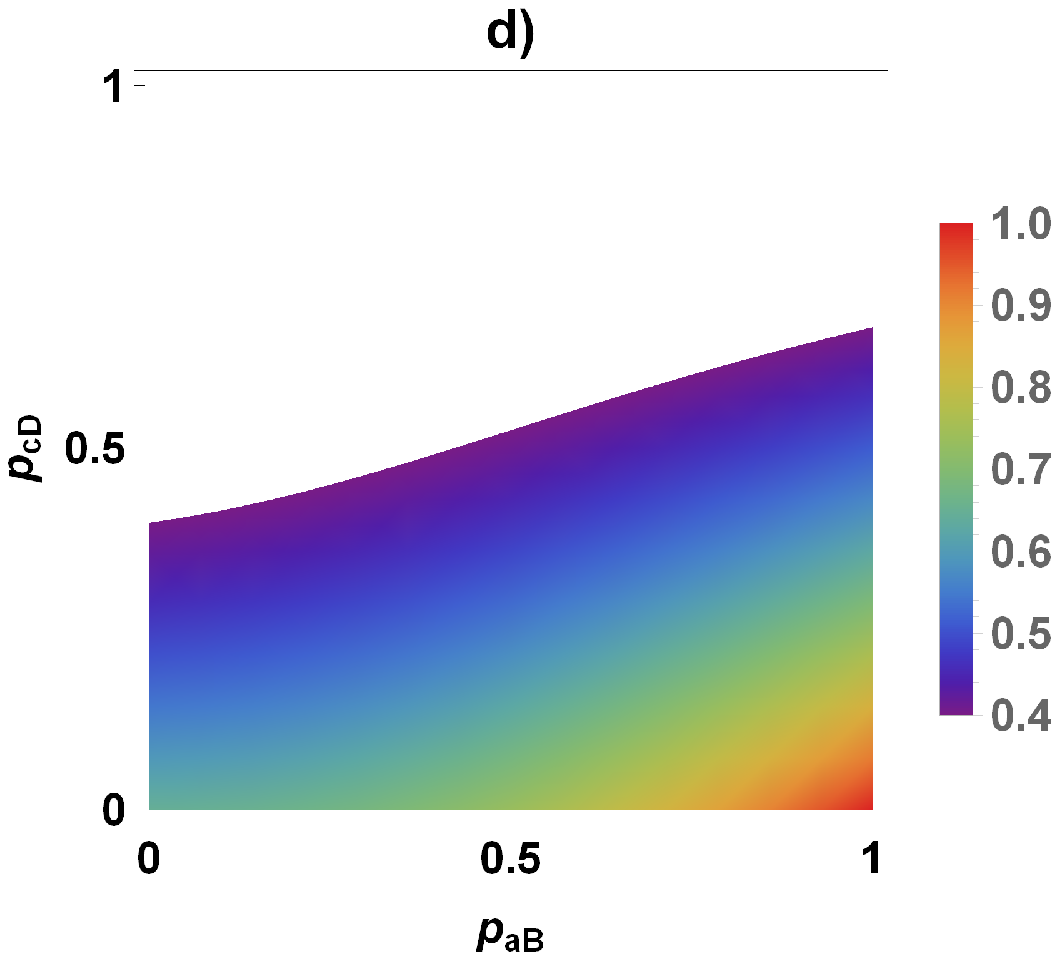}
\end{center}\vspace{-0.4cm}
\caption{Phase diagrams of the optimal averaged fidelities a) $\left\langle F_{BB}\right\rangle_{opt}$, b) $\left\langle F_{BP}\right\rangle_{opt}$, c) $\left\langle F_{BA}\right\rangle_{opt}$, and d) $\left\langle F_{BD}\right\rangle_{opt}$ in the $p_{aB}-p_{c\gamma}$ spaces in case all qubits subjected to bit-flip noise are applied to the Pauli operator $X$ before being sent through noisy environments. Colors illustrate the quantum values of $\left\langle F_{B\gamma}\right\rangle_{opt}$ and white background shows the classical domain.}
\label{fgBL}
\end{figure}\\
Next, let's consider $\alpha = P$ and $\gamma \in \{B,P,A,D\}$. The optimal averaged fidelities $\left\langle F_{P\gamma}\right\rangle_{opt}$ are achieved with the following values of $\theta^{(P\gamma)}_{opt}$ and $\xi^{(P\gamma)}_{opt}$
\begin{eqnarray}
\theta_{opt}^{(PB)}=\xi^{(PB)}_{opt}=\theta_{opt}^{(PD)}=\xi^{(PD)}_{opt}=\xi^{(PA)}_{opt}=
\cases{
\pi/4 &for $p_{aP}<1/2$,\\
-\pi/4 &for $p_{aP}>1/2$,\\}\\
\theta_{opt}^{(PP)}=\xi_{opt}^{(PP)}=
\cases{
\pi/4 &for $(1-2p_{aP})(1-2p_{cP})>0$,\\
-\pi/4 &for $(1-2p_{aP})(1-2p_{cP})<0$\\}
\end{eqnarray}
and
\begin{eqnarray}\label{thetaPA}
\theta^{(PA)}_{opt}=\frac{1}{2}\arctan\frac{2 \left(1-2 p_{{aP}}\right) \sqrt{1-p_{{cA}}}}{p_{{cA}}}
\end{eqnarray} 
with $\sin{2\theta^{(PA)}_{opt}}>0$ and $\cos{2\theta^{(PA)}_{opt}>0}$ for $p_{aP}<1/2$ or  $\sin{2\theta^{(PA)}_{opt}}<0$ and $\cos{2\theta^{(PA)}_{opt}>0}$ for $p_{aP}>1/2$. The detailed optimal expressions of $\left\langle F_{P\gamma}\right\rangle_{opt}$ are attached in Eqs. (\ref{A1}) - (\ref{A4}) in Appendix A.
\begin{figure}[ht!]
\begin{center}
\includegraphics[scale=0.55]{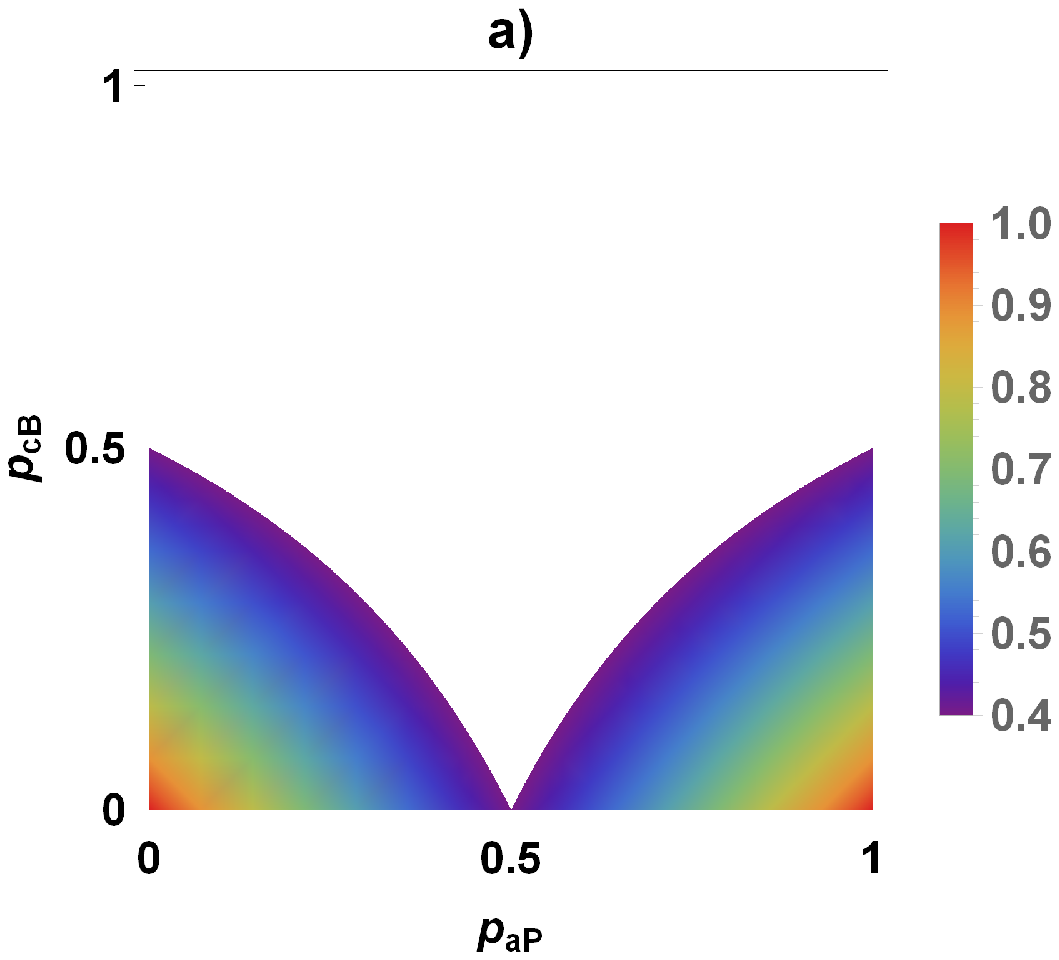}\quad\quad
\includegraphics[scale=0.55]{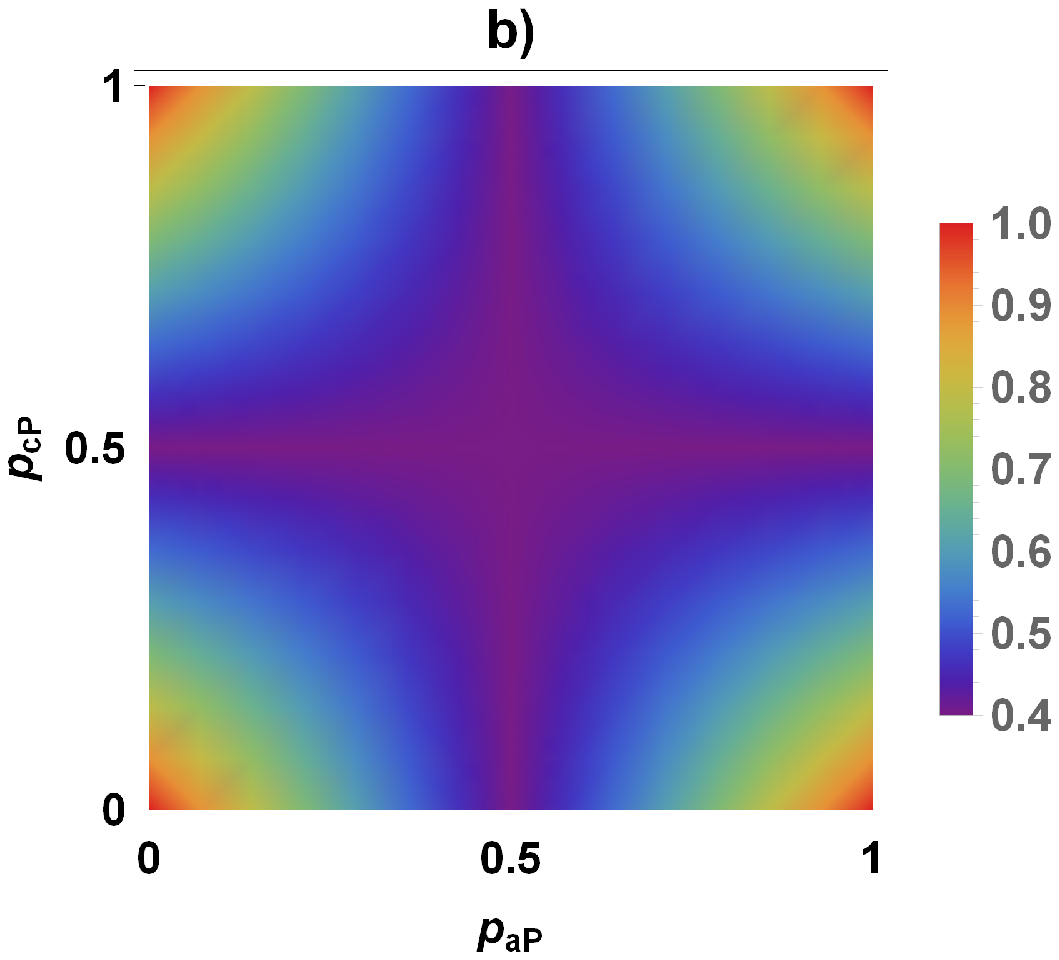}\vspace{0.2cm}
\includegraphics[scale=0.55]{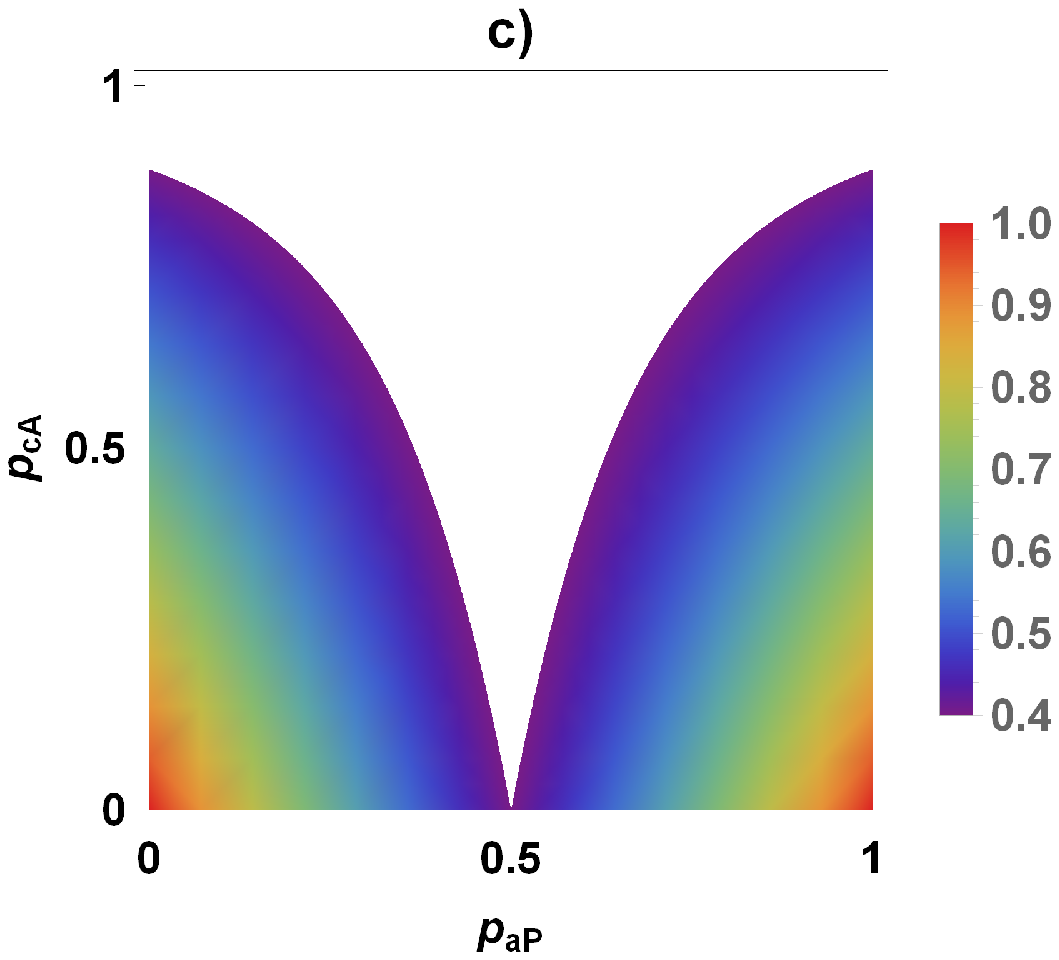}\quad\quad
\includegraphics[scale=0.55]{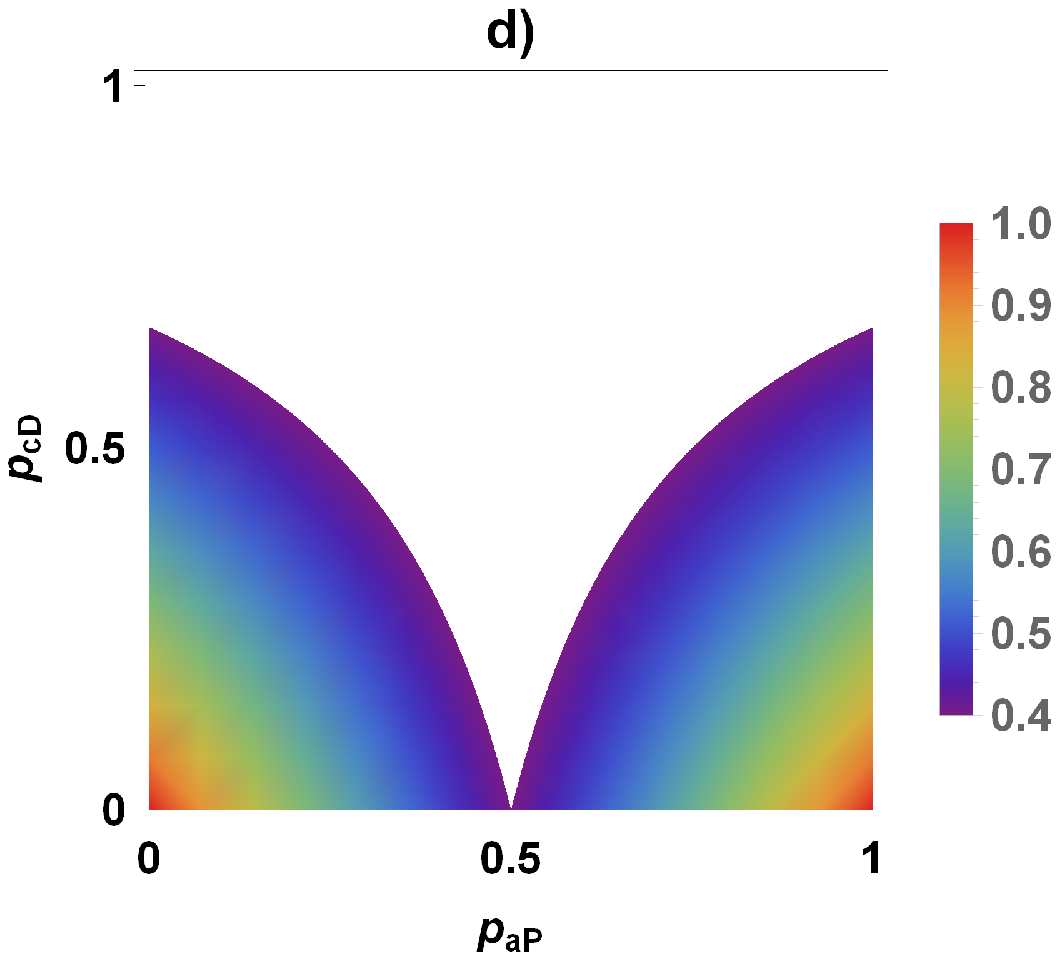}
\end{center}\vspace{-0.4cm}
\caption{Phase diagrams of the optimal averaged fidelities a) $\left\langle F_{PB}\right\rangle_{opt}$, b) $\left\langle F_{PP}\right\rangle_{opt}$, c) $\left\langle F_{PA}\right\rangle_{opt}$, and d) $\left\langle F_{PD}\right\rangle_{opt}$ in the $p_{aB}-p_{c\gamma}$ spaces. Colors illustrate the quantum values of $\left\langle F_{P\gamma}\right\rangle_{opt}$ and white background shows the classical domain.}
\label{fgP}
\end{figure}\\
The quantum domains of $\left\langle F_{P\gamma}\right\rangle_{opt}$ are present in Fig. (\ref{fgP}). It can be easily seen that the useful regions in Figs. 3a, 3c, and 3d have similar patterns with the symmetry with respect to the segment $p_{aP}=1/2$. However, the quantum area in case of $\left\langle F_{PA}\right\rangle_{opt}$ is greater than those in case of either $\left\langle F_{PB}\right\rangle_{opt}$ or $\left\langle F_{PD}\right\rangle_{opt}$. The last one, Fig. 3b, shows that the quantum area is symmetric with respect to not only the segment $p_{aP}=1/2$ but also the one $p_{cP}=1/2$ and spreads over the full parameter ranges. Therefore, this is an unexpected result since no matter how strong noises are the protocol always remains its usefulness. The reason for those symmetries, however, is similar to what explained in Fig. 1b and the quantum domain of $\left\langle F_{PB}\right\rangle_{opt}$ can be found in a bigger range of $p_{cB}$ by employing the same scheme whose result is demonstrated in Fig. (\ref{fgBL}). It's again noteworthy that the value of $\theta_{opt}^{(PA)}$  in Eq. (\ref{thetaPA}) is not required to be equal to $\pi/4$, which results in the best JRSP performed with less entanglement. \\ 
Then, address the scenario in which $\alpha=A$ and $\gamma \in \{B,P,A,D\}$. The expressions of $\theta_{opt}^{(A\gamma)}$ and $\xi_{opt}^{(A\gamma)}$ reads
\begin{eqnarray}
  \xi_{opt}^{(AB)}=\xi_{opt}^{(AA)}=\xi_{opt}^{(AD)}=\frac{\pi}{4},\\
\fl  \theta_{opt}^{(AB)}=\frac{1}{2}\arctan \nonumber\\
\fl \frac{\left(1-p_{{cB}}\right)^2\Big\{\sqrt{1-p_{{aA}}} \big[(\pi ^2-16) p_{{aA}}+32\big]+32(1-p_{aA})\Big\}}{p_{{aA}} \Big\{\left(16-\pi ^2\right) \sqrt{1-p_{{aA}}} \left(1-p_{{cB}}\right)^2-8 \left(2 p_{{cB}}-1\right) \big[2 \left(p_{{aA}}-1\right) p_{{cB}}-p_{{aA}}+2\big]\Big\}}
\end{eqnarray}
with $\sin{2\theta_{opt}^{(AB)}}>0$ and $\cos{2\theta_{opt}^{(AB)}}>0$ for $(p_{aA},p_{cB})$ satisfying the inequality (obviously $0\leq p_{aA},p_{cB}\leq 1$)
\begin{equation}
\fl p_{{aA}} \Big\{\left(16-\pi ^2\right) \sqrt{1-p_{{aA}}} \left(1-p_{{cB}}\right)^2-8 \left(2 p_{{cB}}-1\right) \big[2 \left(p_{{aA}}-1\right) p_{{cB}}-p_{{aA}}+2\big]\Big\}>0
\end{equation}
or with $\sin{2\theta_{opt}^{(AB)}}>0$ and $\cos{2\theta_{opt}^{(AB)}}<0$ for $(p_{aA},p_{cB})$ satisfying the inequality
\begin{equation}
\fl p_{{aA}} \Big\{\left(16-\pi ^2\right) \sqrt{1-p_{{aA}}} \left(1-p_{{cB}}\right)^2-8 \left(2 p_{{cB}}-1\right) \big[2 \left(p_{{aA}}-1\right) p_{{cB}}-p_{{aA}}+2\big]\Big\}<0,
\end{equation}
\begin{equation}
 \theta_{opt}^{(AA)}=\frac{1}{2}\arctan\frac{M_{AA}}{N_{AA}}
\end{equation}
with $\sin2\theta_{opt}^{(AA)}>0$ and $\cos2\theta_{opt}^{(AA)}$ for any $0\leq p_{aA},p_{cA}\leq 1$ and
\begin{eqnarray}
\fl M_{AA}=\frac{1}{160} \Big\{32\left(p_{{aA}}-1\right) \left(p_{{cA}}-1\right)+\sqrt{\left(p_{{aA}}-1\right) \left(p_{{cA}}-1\right)} \big[32-(\pi ^2-16) p_{{aA}} \left(p_{{cA}}-1\right) \nonumber \\
\fl -16 p_{{cA}}\big]\Big\}\label{MAA},\\
\fl N_{AA}=\frac{1}{160} \Big\{\sqrt{\left(p_{{aA}}-1\right) \left(p_{{cA}}-1\right)} \big[(\pi ^2-16) p_{{aA}} \left(p_{{cA}}-1\right)+16 p_{{cA}}\big]+8 \big[p_{{aA}} \left(1-2 p_{{cA}}\right) \nonumber \\
\fl +p_{{cA}}\big] \big[p_{{aA}} \left(2 p_{{cA}}-1\right)+ 2-p_{{cA}}\big]\Big\}\label{NAA},
\end{eqnarray}
\begin{eqnarray}
\fl \theta_{opt}^{(AD)}=\frac{1}{2}\arctan \frac{\sqrt{1-p_{{aA}}} \big[(\pi ^2-16) p_{{aA}}+32\big] \left(2-p_{{cD}}\right)+64 \left(p_{{aA}}-1\right) \left(p_{{cD}}-1\right)}{p_{{aA}} \Big\{\big[\left(16-\pi ^2\right) \sqrt{1-p_{{aA}}}+16\big] \left(2-p_{{cD}}\right)+16 p_{{aA}} \left(p_{{cD}}-1\right)\Big\}}
\end{eqnarray}
with $\sin 2\theta_{opt}^{(AD)}>0$ and $\cos 2\theta_{opt}^{(AD)}>0$ for any $0 < p_{aA}\leq1$ and $0\leq p_{cD}<1$,
\begin{eqnarray}
\xi_{opt}^{(AP)}=\cases{\pi/4 &for $p_{cP}<1/2$,\\
-\pi/4 &for $p_{cP}>1/2$\\}
\end{eqnarray}
and
\begin{eqnarray}
\fl \theta_{opt}^{(AP)}=\frac{1}{2}\arctan \nonumber \\
\fl \frac{\left(1-2 p_{{cP}}\right) \Big\{32 \left(p_{{aA}}-1\right) \left(2 p_{{cP}}-1\right)+\sqrt{1-p_{{aA}}} \left[(\pi ^2-16) p_{{aA}}+32\right]\Big\}}{p_{{aA}} \left[\left(\pi ^2-16\right) \sqrt{1-p_{{aA}}} \left(2 p_{{cP}}-1\right)-8 p_{{aA}}+16\right]}
\end{eqnarray}
with $\sin2\theta_{opt}^{(AP)}>0$ and $\cos2\theta_{opt}^{(AP)}>0$ for $p_{cP}<1/2$ or with $\sin2\theta_{opt}^{(AP)}<0$ and $\cos2\theta_{opt}^{(AP)}>0$ for $p_{cP}>1/2$.
The detailed optimal expressions of $\left\langle F_{A\gamma}\right\rangle_{opt}$ are attached in Eqs. (\ref{A5}) - (\ref{A8}) in Appendix A.
\begin{figure}[ht!]
\begin{center}
\includegraphics[scale=0.55]{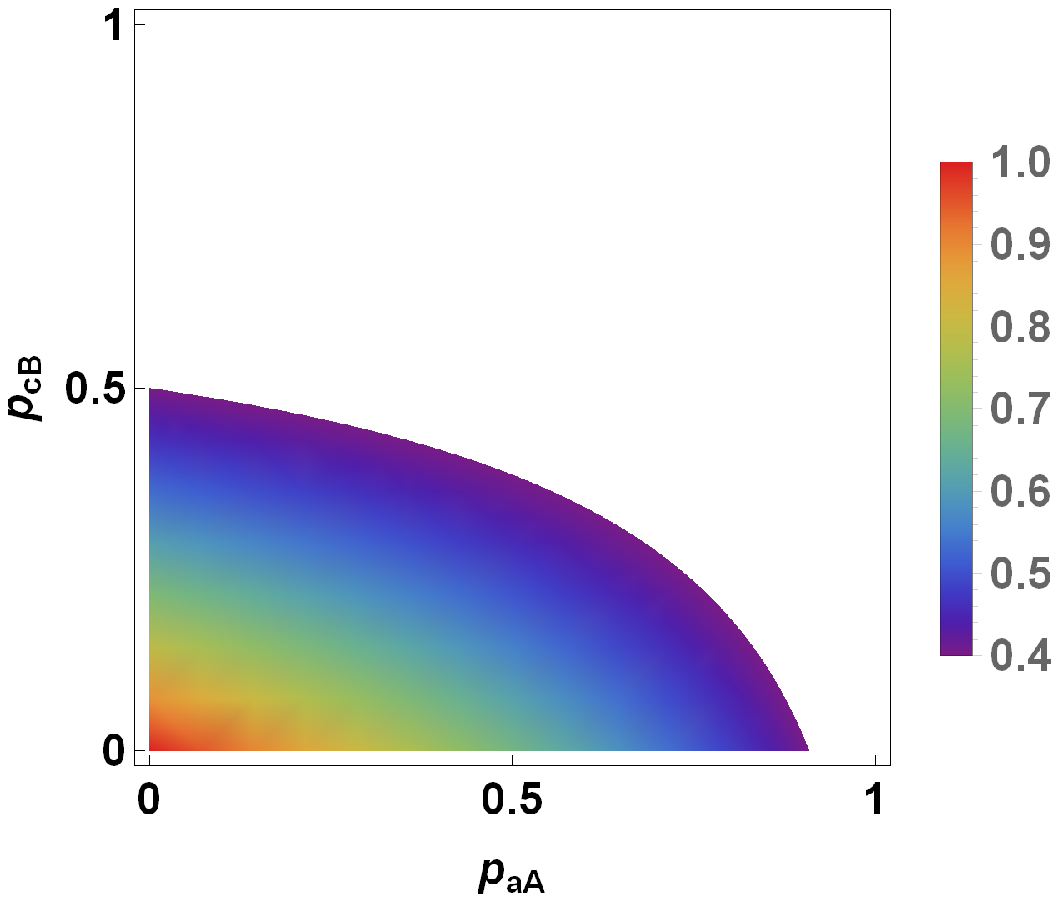}\quad\quad
\includegraphics[scale=0.55]{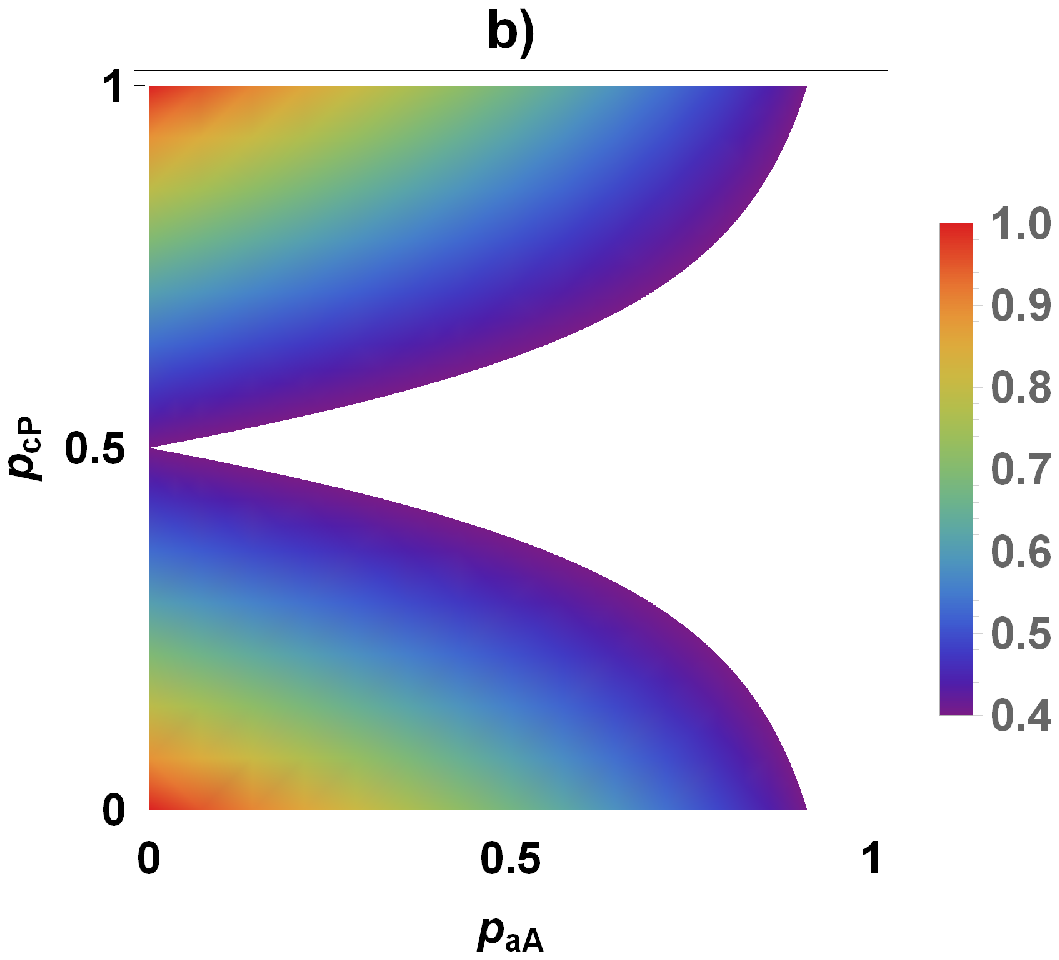}\vspace{0.2cm}
\includegraphics[scale=0.55]{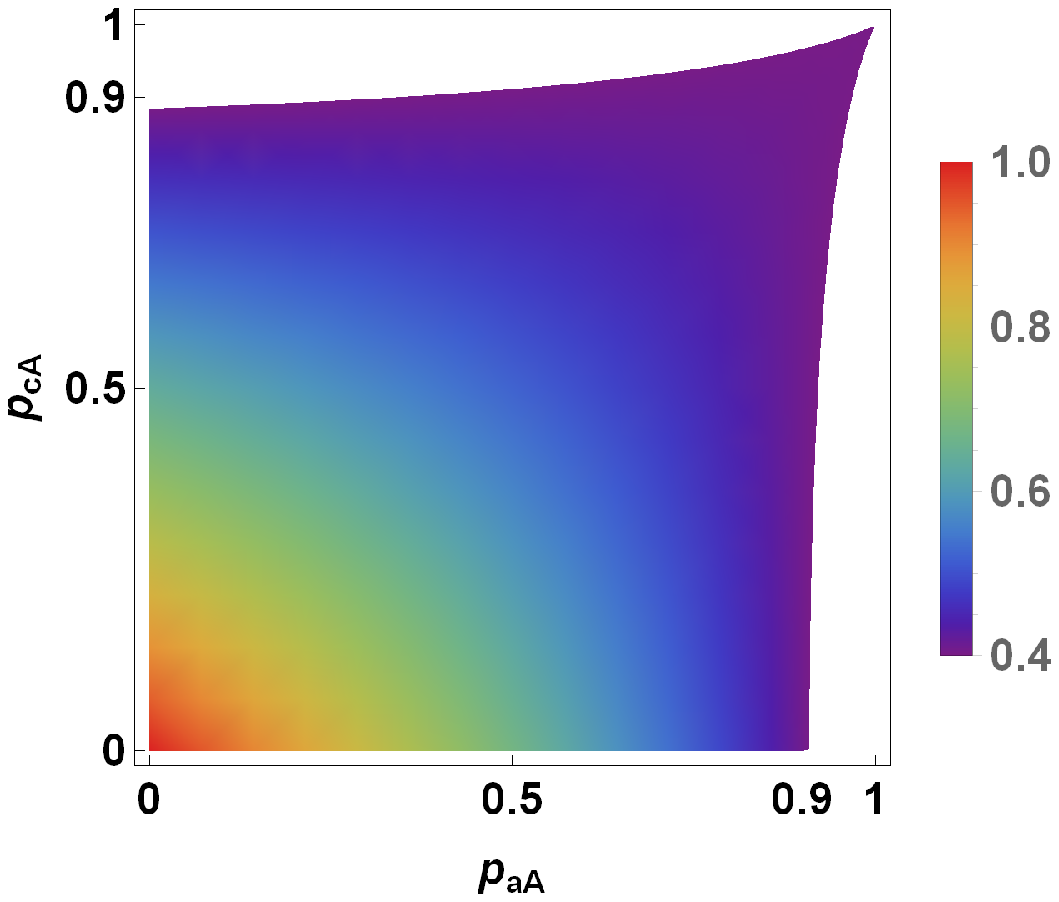}\quad\quad
\includegraphics[scale=0.55]{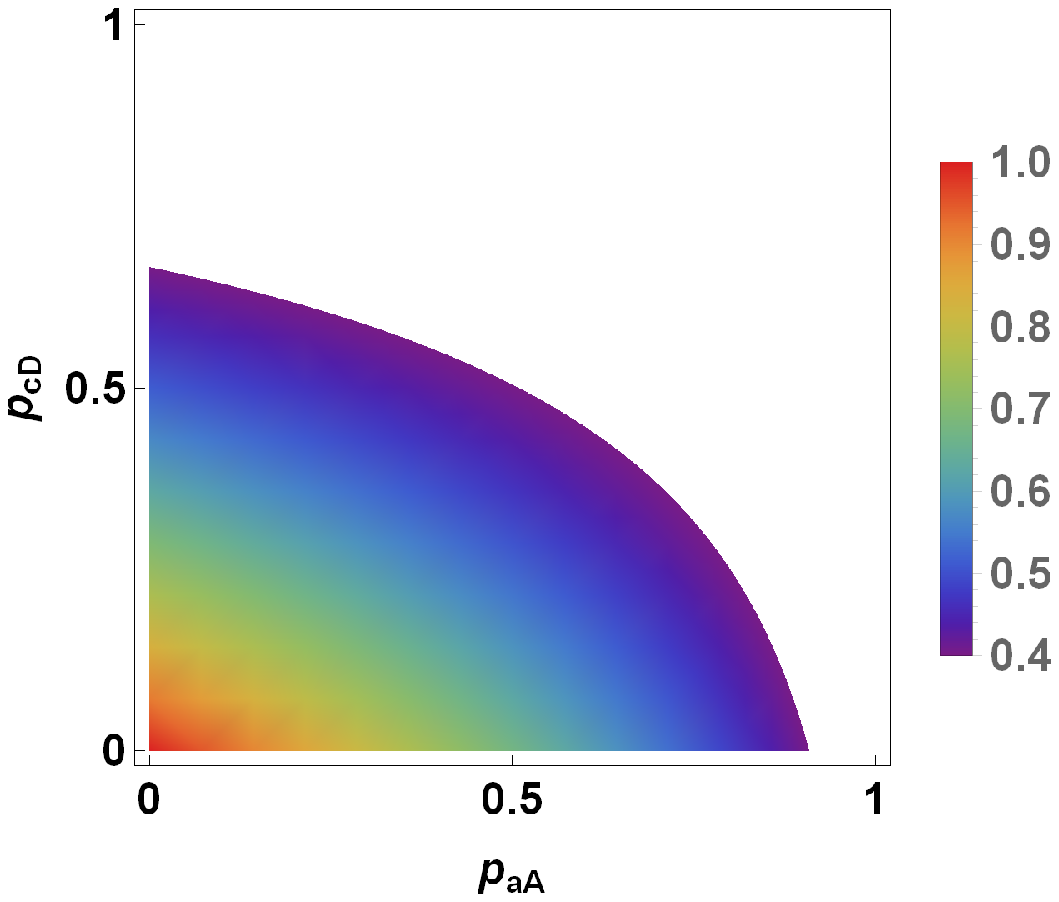}
\end{center}\vspace{-0.4cm}
\caption{Phase diagrams of the optimal averaged fidelities a) $\left\langle F_{AB}\right\rangle_{opt}$, b) $\left\langle F_{AP}\right\rangle_{opt}$, c) $\left\langle F_{AA}\right\rangle_{opt}$, and d) $\left\langle F_{AD}\right\rangle_{opt}$ in the $p_{aA}-p_{c\gamma}$ spaces. Colors illustrate the quantum values of $\left\langle F_{A\gamma}\right\rangle_{opt}$ and white background shows the classical domain.}
\label{fgA}
\end{figure}\\
In Fig. (\ref{fgA}), as the quantum domain spreads over almost values of noise strength parameters, the noise pair $(\alpha,\gamma)=(A,A)$ exhibits better quality than the other three. Furthermore, while $\left\langle F_{A\gamma}\right\rangle_{opt}\,(\gamma \in \{B,D\})$ decrease with any rises in noise parameters, there is a region in which a quantum value of $\left\langle F_{AA}\right\rangle_{opt}$ is found even with larger noise parameters. To clarify more detailed about such result, the plot in Fig. (\ref{figAA}) comparing the optimal averaged fidelity in case $p_{cA}=0$ and $p_{aP}$ is rewritten as $p_{A}$, noted as $\left\langle F_{A0}\right\rangle_{opt}$, the second one in case $p_{aA}=p_{cA}=p_{A}$, noted as $\langle \tilde{F}_{AA}\rangle_{opt}$ and the averaged fidelity in case $\theta=\xi=\pi/4$ and $p_{aA}=p_{cP}=p_{A}$, noted as $\left\langle F_{AA}\right\rangle$, is present. From the plot, one can check that $\langle \tilde{F}_{AA}\rangle_{opt} <\left\langle F_{A0}\right\rangle_{opt}$ with almost the range of $p_{A}$, but with large enough $p_{A}$, $\langle \tilde{F}_{AA}\rangle_{opt} >\left\langle F_{A0}\right\rangle_{opt}$, which means adding more amplitude-damping noise can improve the fidelity of the protocol. If the initial $\theta$ and $\xi$ are chosen as $\pi/4$, the averaged fidelity, interestingly, behaves in such a way that larger enough $p_{A}$ can increase $\left\langle F_{AA}\right\rangle$. However, for any $0< p_{A}\leq 1$, $\left\langle F_{AA}\right\rangle < \langle \tilde{F}_{AA}\rangle_{opt}$ and $\langle \tilde{F}_{AA}\rangle_{opt} \geq 2/5$, expressing the useful role of $\theta_{opt}^{(AA)}$. Physical mechanism for such adding more noise of amplitude-damping leading to a larger fidelity is possibly the same to what showed in Ref. \cite{i36} but for quantum teleportation, local dissipative environment can enhance the quality of the protocol.  The effect of these dissipative environments is represented by virtue of trace-preserving and completely positive maps. Moreover, the above result is quietly in accordance with the ones obtained in Refs. \cite{i35, i37} which found that in a specific domain of noisy strengths, the quantum teleportation fidelity in case of two qubits simultaneously subjected to amplitude-damping noise is higher than that in case of only one qubit affected by that noise.     
\begin{figure}[ht!]
\begin{center}
\includegraphics[scale=1]{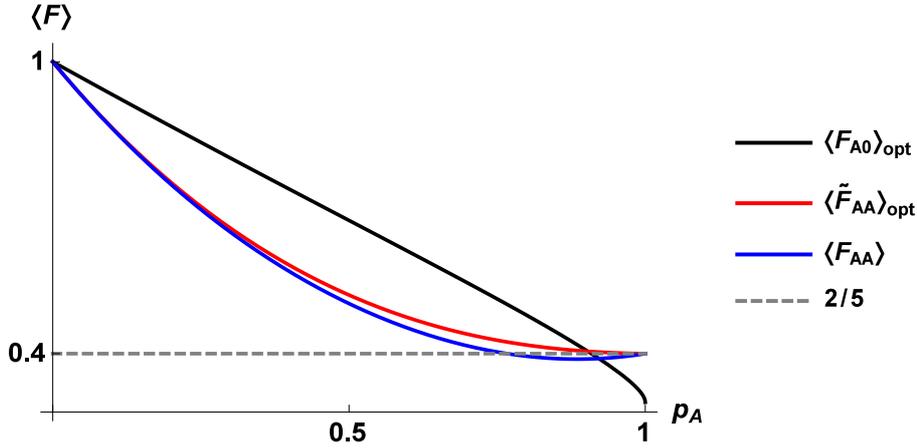}
\end{center}
\vspace{-0.3 cm}
\caption{The plot compares $\left\langle F_{A0}\right\rangle_{opt},\, \langle \tilde{F}_{AA}\rangle_{opt},$ and $\left\langle F_{AA}\right\rangle$ as the functions of $p_{A}$.}
\label{figAA}
\end{figure}\\
Finally, the scenario in which $\alpha=D$ and $\gamma\in\{B,P,A,D\}$ is took into account and its results are
\begin{eqnarray}
 \theta_{opt}^{(DB)}=\xi_{opt}^{(DB)}=\theta_{opt}^{(DD)}=\xi_{opt}^{(DD)}=\xi^{(DA)}_{opt}=\frac{\pi}{4},\\
 \theta_{opt}^{(DP)}=\xi_{opt}^{(DP)}=
\cases{\pi/4 &for $p_{cP}<1/2$,\\
-\pi/4 &for $p_{cP}>1/2$
}
\end{eqnarray}
and 
\begin{eqnarray}
\fl \theta_{opt}^{(DA)}=\frac{1}{2}\arctan\frac{\left[(\pi ^2-16) p_{{aD}}+32\right] \sqrt{1-p_{{cA}}} \left(2-p_{{cA}}\right)+64 \left(p_{{aD}}-1\right) \left(p_{{cA}}-1\right)}{p_{{cA}} \Big\{\big[(\pi ^2-16) p_{{aD}}+32\big] \sqrt{1-p_{{cA}}}+16 \big[p_{{aD}} \left(p_{{cA}}-1\right)-p_{{cA}}+2\big]\Big\}}
\end{eqnarray}
with $\sin{2\theta_{opt}^{(DA)}}>0$ and $\cos2\theta_{opt}^{(DA)}>0$ for any $0\leq p_{aD}<1$ and $0< p_{cA}\leq 1$. The detailed optimal expressions of $\left\langle F_{D\gamma}\right\rangle_{opt}$ are attached in Eqs. (\ref{A9}) - (\ref{A12}) in Appendix A.
\begin{figure}[ht!]
\begin{center}
\includegraphics[scale=0.55]{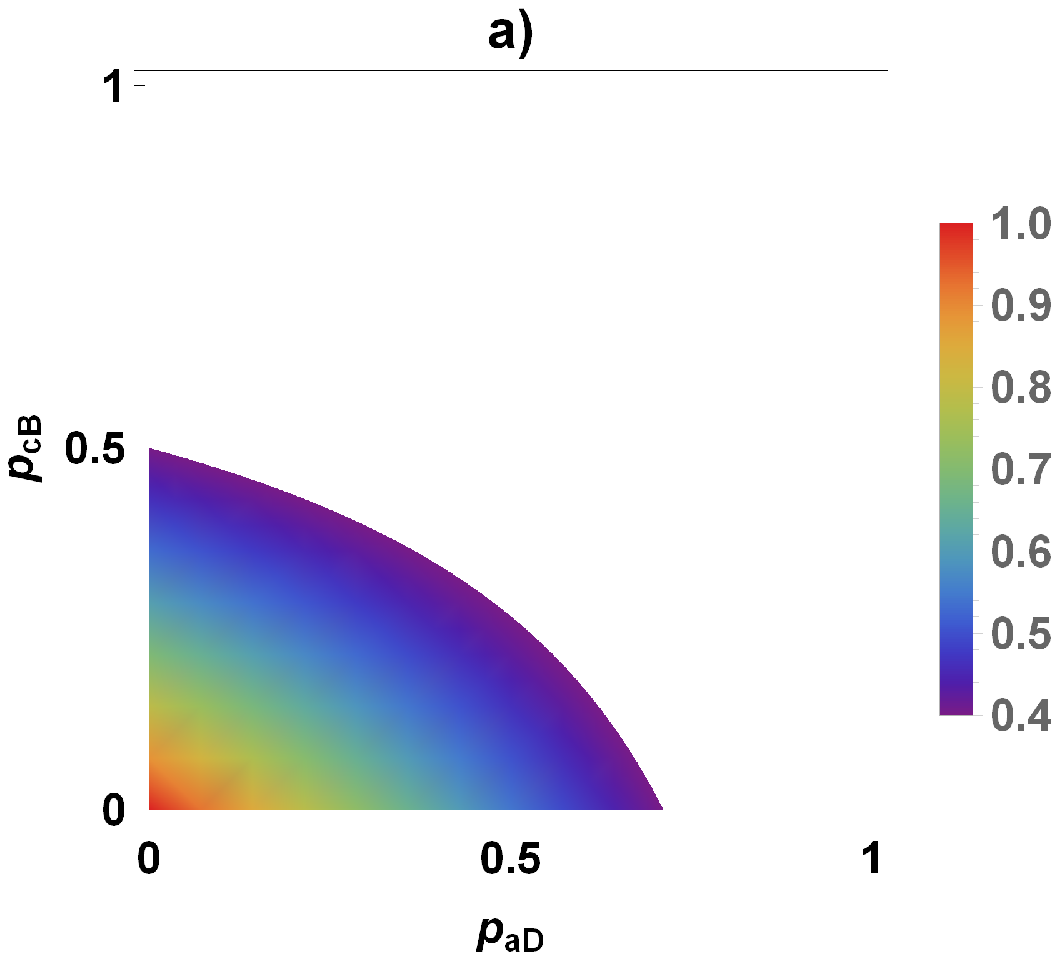}\quad\quad
\includegraphics[scale=0.55]{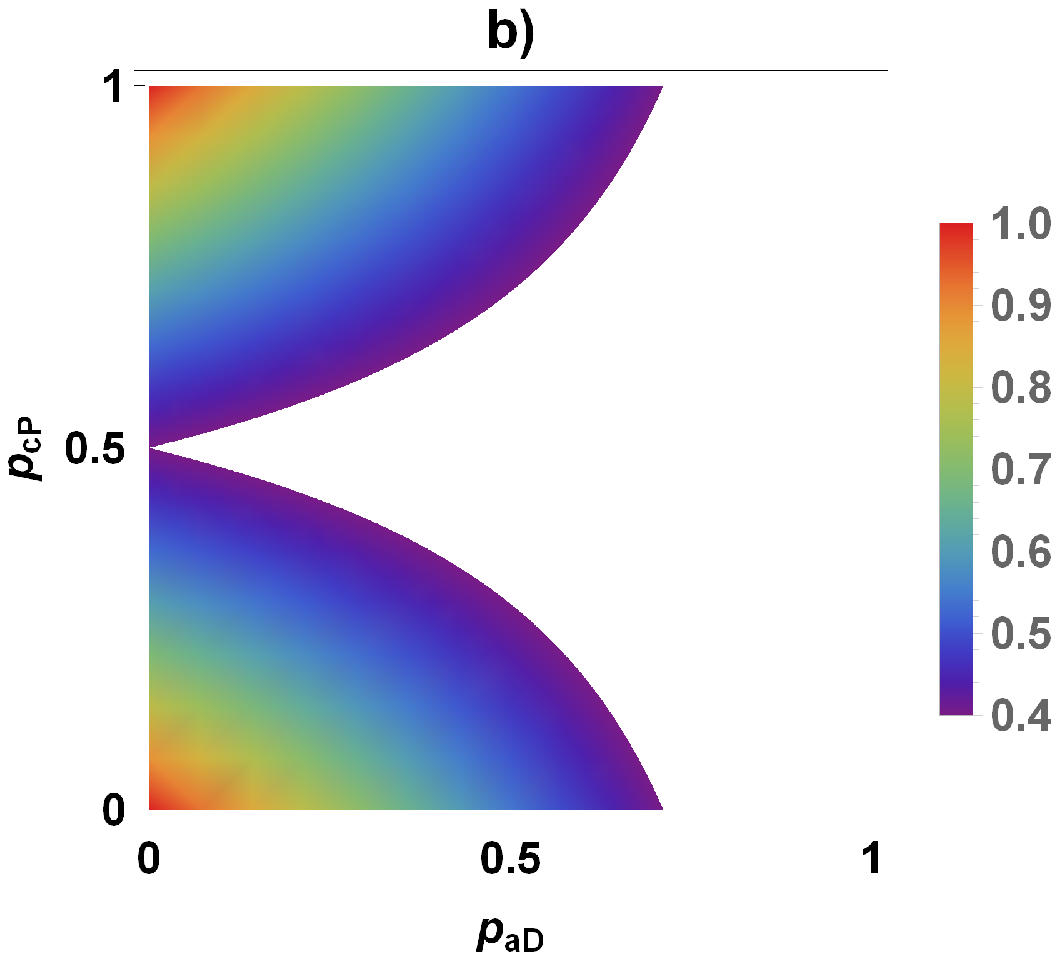}\vspace{0.2cm}
\includegraphics[scale=0.55]{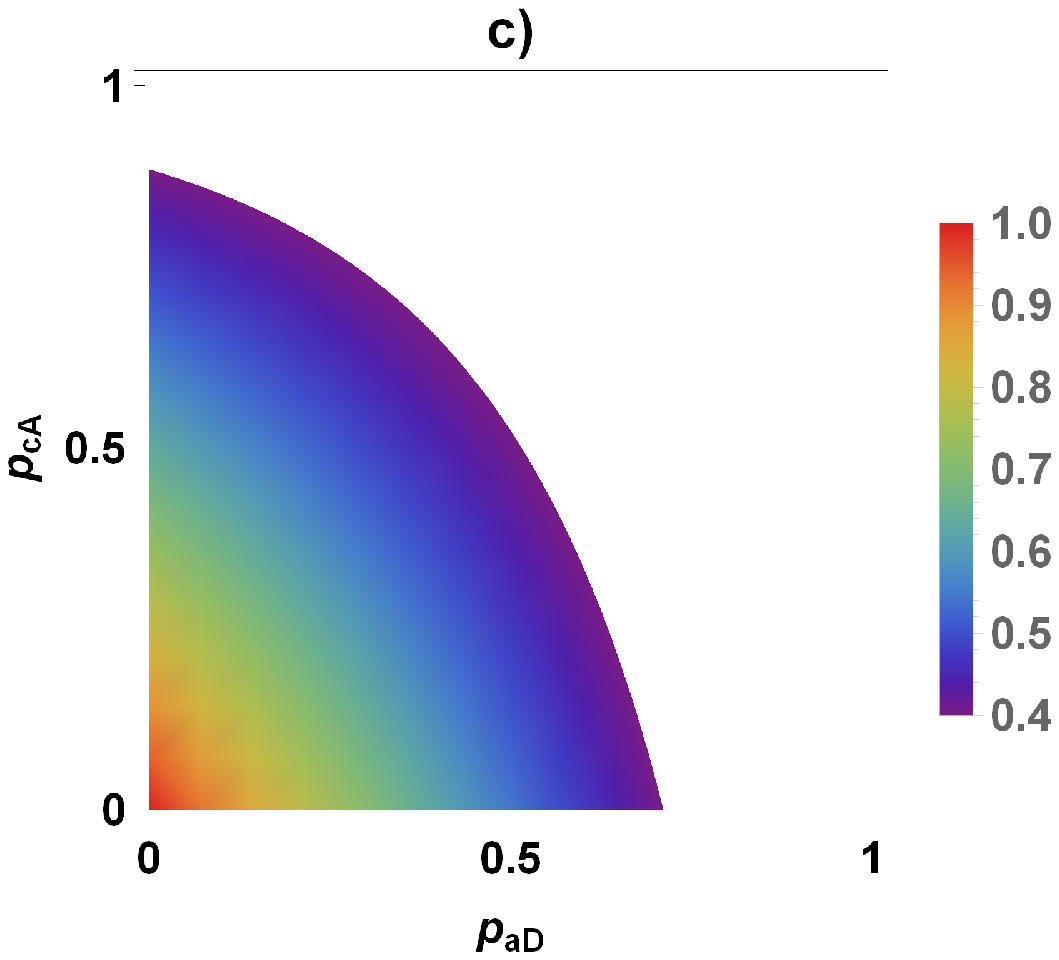}\quad\quad
\includegraphics[scale=0.55]{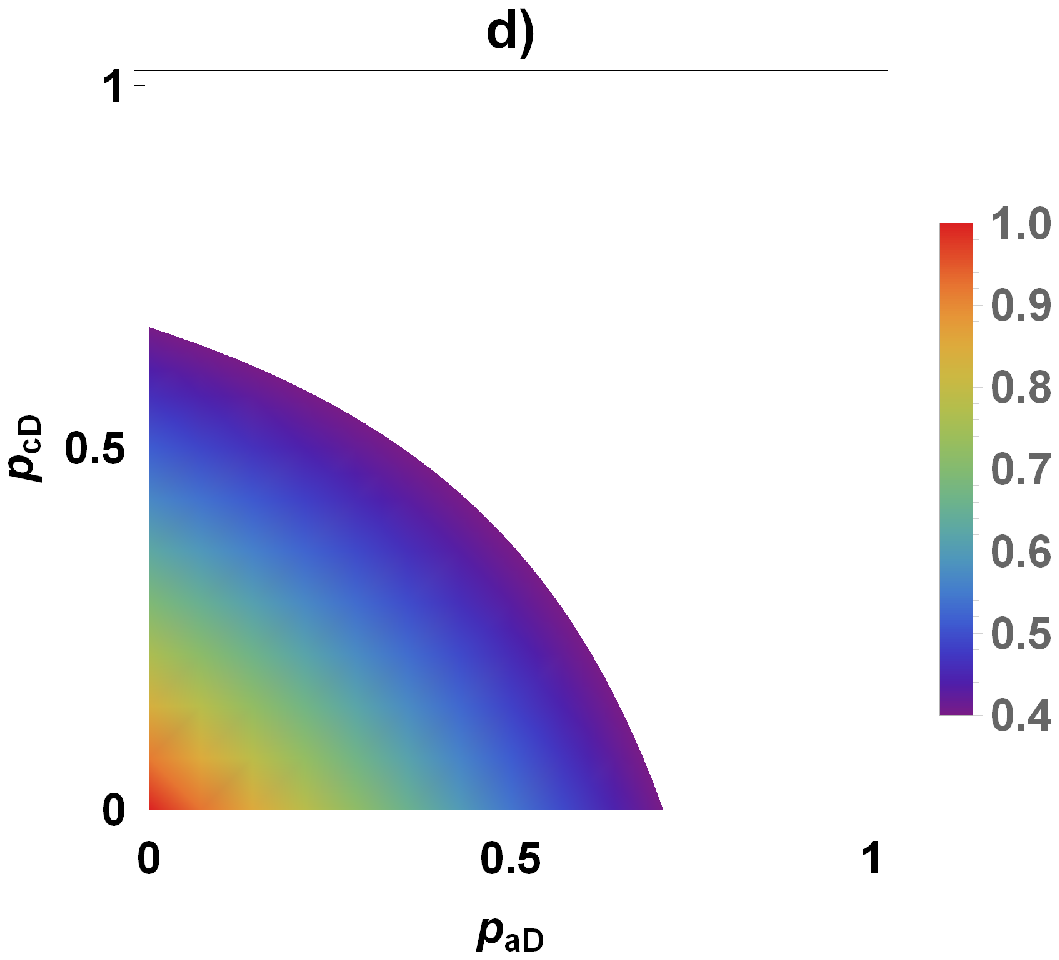}
\end{center}\vspace{-0.4cm}
\caption{Phase diagram of the optimal averaged fidelities a) $\left\langle F_{DB}\right\rangle_{opt}$, b) $\left\langle F_{DP}\right\rangle_{opt}$, c) $\left\langle F_{DA}\right\rangle_{opt}$, and d) $\left\langle F_{DD}\right\rangle_{opt}$ in the $p_{aD}-p_{c\gamma}$ spaces. Colors illustrate the quantum values of $\left\langle F_{D\gamma}\right\rangle_{opt}$ and white background shows the classical domain.}
\label{fgD}
\end{figure}\\
Fig. (\ref{fgD}) shows no surprise in the useful regions of $\left\langle F_{D\gamma}\right\rangle_{opt}$. Similar to what analyzed in Figs. (\ref{fgB}), (\ref{fgP}) and (\ref{fgA}), the quantum area of $\left\langle F_{DA}\right\rangle_{opt}$ keeps superior to those of $\left\langle F_{D\gamma}\right\rangle_{opt}\left(\gamma \in \{B,D\}\right)$ and there is a symmetry at the change of $\left\langle F_{DP}\right\rangle_{opt}$.
\section{Conclusion}
We have studied the quality of the joint remote state preparation of a two-qubit state under the influence of four types of noise, the bit-flip, phase-flip, amplitude-damping and depolarizing. In order to describe the action of noises on qubits superoperators being in the operator-sum of Kraus operators are exploited. It is supposed that independently, two qubits of not only the first sender but also the receiver suffer the same type of noise. The corresponding optimal averaged fidelities were derived and analyzed in phase-space diagrams to clarify the domain of noise strength parameters in which the efficiency of the protocol exhibits quantumly. The symmetrical behavior of the optimal averaged fidelity of the protocol subjected to the phase-flip noise has been basically explained. Besides, the fidelity under the influence of bit-flip noise is also optimized in the sense that Bob, who produces the quantum channel, applies the Pauli operator $X$ to qubits which are going to be sent through bit-flip environments in case he knows the parameters are large. Essentially, depending on the range of noisy parameter, the choices of $\theta$ and $\xi$ in case of the phase-flip noise or applying the Pauli operator $X$ in case of bit-flip noise only transforms the initial state of qubits 1, 3 and 5 (2, 4 and 6) from one of the GHZ states into other GHZ states or changes Bob's measurement. Therefore, the optimization of the bit-flip noise as well as that of the phase-flip noise doesn't change the entanglement of the quantum channel and more precisely, is the optimization of the steps of JRSP protocol. In contrast to this, the optimization for JRSP protocol affected by the amplitude-damping noise showed that the value of $\theta^{(A\gamma)}_{opt}$ or $\theta^{(\alpha A)}_{opt}$ is varied with the change of noise parameters and possibly different from $\pi/4$, which in principle makes the entanglement of the quantum channel changed. Remarkably, when qubits 1 and 2 are suffered the amplitude-damping noise, adding another noise acting on qubits 5 and 6 can broaden the area of quantum domain even in considerable noise parameter ranges only if that noise is again amplitude-damping. Such optimization should be interpreted as the optimization which is accomplished through dissipative interactions with noisy environments. From these results, we hope to shed more light on improving the realistic manipulation of JRSP of an arbitrary two-qubit state. 

\ack
This work is supported by the Vietnam National Foundation for Science and Technology Development (NAFOSTED) under project No.103.01-2017.08

\clearpage

\appendix

\section{The detailed expressions of the optimal averaged fidelities}
\begin{eqnarray}
\fl \left\langle F_{PB}\right\rangle _{opt} &=& \frac{2}{5}+\frac{1}{5} \left(p_{{cB}}-2\right) p_{{cB}}+\frac{1}{5} \left(2 p_{{aP}}-1\right)^2 \left(p_{{cB}}-1\right)^2+\frac{2}{5} \left(p_{{cB}}-1\right)^2 \left| 1-2 p_{{aP}}\right|, \nonumber \\
 \fl &&\label{A1}\\
\fl \left\langle F_{PP}\right\rangle_{opt}&=&\frac{2}{5}+\frac{1}{5} \left(2 p_{{aP}}-1\right)^2 \left(2 p_{{cP}}-1\right)^2+\frac{2}{5} \big| \left(2 p_{{aP}}-1\right) \left(2 p_{{cP}}-1\right)\big|,
\label{A2}\\
\fl \left\langle F_{PA}\right\rangle_{opt} &=& \frac{2}{5}+\frac{1}{20} \left(p_{{cA}}-4\right) p_{{cA}}+\frac{1}{10} \sqrt{1-p_{{cA}}} \left(2-p_{{cA}}\right)\left| 1-2 p_{{aP}}\right|+\frac{1}{20}\Big\{\Big\{2\left(1-2 p_{{aP}}\right) \nonumber \\
\fl && \times  \big[2 \left(1-p_{{cA}}\right) \left| 1-2 p_{{aP}}\right| +\sqrt{1-p_{{cA}}} \left(2-p_{{cA}}\right)\big]\Big\}^2+\Big\{p_{{cA}} \big[2 \sqrt{1-p_{{cA}}} | 1 \nonumber \\
\fl &&-2 p_{{aP}}| +2  - p_{cA} \big] \Big\}^2\Big\}^{1/2}
\label{A3},\\
\fl \left\langle F_{PD}\right\rangle_{opt} &=& \frac{2}{5}+
\frac{1}{20} \left(p_{{cD}}-4\right) p_{{cD}}+
\frac{1}{5} \left(2 p_{{aP}}-1\right)^2 \left(p_{{cD}}-1\right)^2+
\frac{1}{5} \left(p_{{cD}}-2\right) \left(p_{{cD}}-1\right) | 1 \nonumber \\
\fl && -2p_{{aP}}|
\label{A4},\\
\fl \left\langle F_{AB}\right\rangle_{opt} &=& \frac{2}{5}+\frac{1}{160} \Big\{\sqrt{1-p_{{aA}}} \big[(\pi ^2-16) p_{{aA}}+32\big] \left(p_{{cB}}-1\right){}^2+8 \big[2 \left(p_{{aA}}-1\right) p_{{cB}}-p_{{aA}}\big] \nonumber \\
\fl && \times \big[2 \left(p_{{aA}}-1\right) p_{{cB}}-p_{{aA}}+ 4\big]\Big\}+\frac{1}{160}\Big\{ \left(p_{{cB}}-1\right)^4\big[32-32 p_{{aA}}+\sqrt{1-p_{{aA}}} \nonumber\\
\fl && \times  (\pi ^2 p_{{aA}} -16 p_{{aA}} +32)\big]^2+p_{{aA}}^2 \Big\{\left(16-\pi ^2\right) \sqrt{1-p_{{aA}}} \left(p_{{cB}}-1\right)^2-8 (2 p_{{cB}} \nonumber \\
\fl && -1)  \big[2 \left(p_{{aA}}-1\right) p_{{cB}} -p_{{aA}} +2\big]\Big\}^2\Big\}^{1/2} \label{A5},\\
\fl \left\langle F_{AP}\right\rangle _{opt} &=& \frac{2}{5}+\frac{1}{20} \left(p_{{aA}}-4\right) p_{{aA}}+\frac{1}{160} \sqrt{1-p_{{aA}}} \big[(\pi ^2-16) p_{{aA}}+32\big] \left| 1-2 p_{{cP}}\right|+\frac{1}{160} \nonumber \\
\fl && \times \Big\{(1-2 p_{{cP}})^2\Big\{\sqrt{1-p_{{aA}}} \big[(\pi ^2-16) p_{{aA}}+32\big]+32 \left(1-p_{{aA}}\right) \left| 1-2 p_{{cP}}\right|
\Big\}^2 \nonumber\\
\fl && +p_{{aA}}^2 \big[(16-\pi ^2) \sqrt{1-p_{{aA}}} \left| 1-2 p_{{cP}}\right| +8 (2-p_{{aA}})
\big]^2 \Big\}\label{A6},\\
\fl \left\langle F_{AA}\right\rangle_{opt} &=& \frac{2}{5}+\frac{1}{160} \Big\{\sqrt{\left(p_{{aA}}-1\right) \left(p_{{cA}}-1\right)} \big[32-(\pi ^2-16) p_{{aA}} \left(p_{{cA}}-1\right)-16 p_{{cA}}\big]+8 \nonumber \\
\fl && \times  \big[p_{{aA}} \left(2 p_{{cA}}-1\right)-p_{{cA}}\big] \big[p_{{aA}} \left(2 p_{{cA}}-1\right)-p_{{cA}}+4\big]\Big\}+\left(M_{AA}^2+N_{AA}^2\right)^{1/2}\label{A7}
\end{eqnarray}
with $ M_{AA}$ and $N_{AA}$ defined as in Eqs. (\ref{MAA}) and (\ref{NAA}),
\begin{eqnarray}
\fl \left\langle F_{AD}\right\rangle_{opt} &=& \frac{2}{5}+\frac{1}{320} \Big\{16 \big[(p_{{aA}} \left(p_{{cD}}-1\right)-p_{{cD}}\big] \big[p_{{aA}} \left(p_{{cD}}-1\right)-p_{{cD}}+4\big]+\sqrt{1-p_{{aA}}} \big[(\pi ^2 \nonumber \\
\fl && -16) p_{{aA}}+32\big] (p_{{cD}}-2) \left(p_{{cD}}-1\right)\Big\}+\frac{1}{320}\Big\{\left(1-p_{{cD}}\right)^2 \Big\{\sqrt{1-p_{{aA}}} \big[(\pi ^2 \nonumber \\
\fl && -16) p_{{aA}} +32\big] (2 -p_{{cD}})+64 \left(p_{{aA}}-1\right) (p_{{cD}}-1)\Big\}^2+p_{{aA}}^2 \left(1-p_{{cD}}\right)^2 \Big\{\big[(16 \nonumber \\
\fl && -\pi ^2) \sqrt{1-p_{{aA}}}+16\big] \left(2-p_{{cD}}\right) +16 p_{{aA}} \left(p_{{cD}}-1\right)\Big\}^2 \Big\}^{1/2}\label{A8},\\
\fl \left\langle F_{DB}\right\rangle_{opt} &=& \frac{2}{5}+\frac{1}{80} \Big\{4 \Big\{p_{{aD}}^2 \big[4 p_{{cB}} \left(3 p_{{cB}}-5\right)+9\big]-4 p_{{aD}} \left(p_{{cB}}-1\right) \left(7 p_{{cB}}-6\right)+4 \big[4 (p_{{cB}} \nonumber\\
\fl && -2)  p_{{cB}}+3\big]\Big\}+\pi ^2 (1-p_{aD}) p_{{aD}} \left(p_{{cB}}-1\right)^2\Big\}\label{A9},\\
\fl \left\langle F_{DP}\right\rangle_{opt} &=& \frac{2}{5}+\frac{1}{80} \Big\{\left(1-p_{{aD}}\right) \big[(\pi ^2-16) p_{{aD}}+32\big] \left| 1-2 p_{{cP}}\right| +16 \left(p_{{aD}}-1\right)^2 \left(1-2 p_{{cP}}\right)^2\nonumber \\
\fl && +4 p_{{aD}}^2-16 p_{{aD}}\Big\}\label{A10},\\
\fl  \left\langle F_{DA}\right\rangle_{opt} &=& \frac{2}{5}+\frac{1}{320} \Big\{\left(1-p_{{aD}}\right) \big[(\pi ^2-16) p_{{aD}}+32\big] \sqrt{1-p_{{cA}}} \left(2-p_{{cA}}\right)+16 \big[p_{{aD}} (p_{{cA}} \nonumber \\
\fl && -1)-p_{{cA}}\big] \big[p_{{aD}} (p_{{cA}}-1)-p_{{cA}}+4\big]\Big\}+\frac{1}{320}\Big\{ \left(1-p_{{aD}}\right)^2 \Big\{64 \left(1-p_{{aD}}\right) (1 \nonumber \\
\fl && -p_{{cA}}) +\big[(\pi ^2 -16) p_{{aD}}+32\big] \sqrt{1-p_{{cA}}} (2-p_{{cA}})\Big\}^2+\left(1-p_{{aD}}\right)^2 p_{{cA}}^2 \Big\{\big[(\pi ^2\nonumber \\
\fl &&-16) p_{{aD}}+32\big] \sqrt{1-p_{{cA}}} +16 \big[p_{{aD}} \left(p_{{cA}}-1\right)-p_{{cA}}+2\big] \Big\}^2 \Big\}^{1/2}\label{A11}
\end{eqnarray}
and 
\begin{eqnarray}
\fl \left\langle F_{DD}\right\rangle_{opt} &=& \frac{2}{5}+\frac{1}{160} \Big\{8 \big[p_{{aD}}^2 \left(p_{{cD}}-1\right) \left(7 p_{{cD}}-9\right)-8 p_{{aD}} \left(p_{{cD}}-1\right) \left(2 p_{{cD}}-3\right)+3 (p_{{cD}} \nonumber \\
\fl && -2) \left(3 p_{{cD}}-2\right)\big]+\pi ^2 (1-p_{aD}) p_{{aD}} \left(2-p_{cD}\right) \left(1-p_{cD}\right)\Big\}\label{A12}.
\end{eqnarray}
\end{document}